\documentclass[twocolumn]{aa}

\usepackage{placeins}
\usepackage{newtxtext}
\usepackage[varg]{txfonts}
\usepackage{graphicx}
\usepackage{natbib}
\usepackage{xcolor}
\usepackage{hyperref}
\usepackage{amsmath}
\usepackage{orcidlink}

\hypersetup{colorlinks=true, linkcolor=blue, citecolor=blue, urlcolor=blue}

\graphicspath{{./}}

\begin{document}

\title{CO(7-6) and [C\,\textsc{i}](2-1) survey in $z>6$ quasars}

\author{
Fuxiang Xu\orcidlink{0000-0003-0754-9795}\inst{1,2,3}\thanks{\email{xufx@stu.pku.edu.cn}}
\and Roberto Decarli\orcidlink{0000-0002-2662-8803}\inst{1}
\and Ran Wang\orcidlink{0000-0003-4956-5742}\inst{2,3}
\and Anna~Elisabetta~Borea\orcidlink{0009-0004-7036-8983}\inst{4,1}
\and Antonio~Pensabene\orcidlink{0000-0001-9815-4953}\inst{5,6,7}
\and Xiaohui~Fan\orcidlink{0000-0003-3310-0131}\inst{8}
\and Dominik~Riechers\orcidlink{0000-0001-9585-1462}\inst{9}
\and Eduardo~Ba\~nados\orcidlink{0000-0002-2931-7824}\inst{10}
\and Axel~Wei{\ss}\orcidlink{0000-0003-4678-3939}\inst{11}
\and Michele~Costa\orcidlink{0009-0009-1622-8636}\inst{4,1}
\and Fabian~Walter\orcidlink{0000-0003-4793-7880}\inst{10,12}
\and Feige~Wang\orcidlink{0000-0002-7633-431X}\inst{13}
\and Jinyi~Yang\orcidlink{0000-0001-5287-4242}\inst{13}
\and Bram~Venemans\orcidlink{0000-0001-9024-8322}\inst{14}
\and Jianan~Li\orcidlink{0000-0002-1815-4839}\inst{15}
\and Emanuele~Paolo~Farina\orcidlink{0000-0002-6822-2254}\inst{16}
}

\institute{
INAF -- Osservatorio di Astrofisica e Scienza dello Spazio, via Gobetti 93/3, I-40129 Bologna, Italy 
\and Department of Astronomy, School of Physics, Peking University, Beijing 100871, China 
\and Kavli Institute for Astronomy and Astrophysics, Peking University, Beijing 100871, China 
\and Department of Physics and Astronomy, University of Bologna, Via Gobetti 93/2, 40129 Bologna, Italy 
\and Dipartimento di Fisica ``G. Occhialini,'' Università degli Studi di Milano-Bicocca, Piazza della Scienza 3, I-20126 Milano, Italy 
\and DTU Space, Technical University of Denmark, Elektrovej 327, DK-2800 Kgs. Lyngby, Denmark 
\and Cosmic Dawn Center (DAWN), Denmark 
\and Steward Observatory, University of Arizona, 933 North Cherry Ave., Tucson, AZ 85721, USA 
\and Institut f\"ur Astrophysik, Universit\"at zu K\"oln, Z\"ulpicher 
Stra{\ss}e 77, D-50937 K\"oln, Germany
\and Max Planck Institut für Astronomie, Königstuhl 17, D-69117 Heidelberg, Germany 
\and Max-Planck-Institut für Radioastronomie, Auf dem Hügel 69, D-53121 Bonn, Germany 
\and National Radio Astronomy Observatory, Pete V. Domenici Array Science Center, P.O. Box O, Socorro, NM 87801, USA 
\and Department of Astronomy, University of Michigan, 1085 S. University Ave., Ann Arbor, MI 48109, USA 
\and Leiden Observatory, Leiden University, P.O. Box 9513, 2300 RA Leiden, The Netherlands 
\and Department of Astronomy, Tsinghua University, Beijing 100084, People's Republic of China 
\and Gemini Observatory, NSF's NOIRLab, 670 North A'ohoku Place, Hilo, HI 96720, USA 
}

\date{Received XX; accepted XX}

\abstract{
High-redshift ($z\gtrsim6$) quasars are signposts of the earliest supermassive black holes and intense star formation, offering key laboratories for black hole--galaxy evolution at cosmic dawn. While far-infrared  studies have revealed large dust reservoirs and strong [C\,\textsc{ii}] emission, the physical condition and molecular gas content of their interstellar medium (ISM) remain uncertain. We present sensitive Atacama Large Millimeter/submillimeter Array Band 3 observations of the redshifted CO(7--6) and [C\,\textsc{i}](2--1) emission lines and the underlying dust continuum in a sample of 18 quasars at $z \sim 6$. We detected CO(7--6) in 15/18, [C\,\textsc{i}](2--1) in 6/18, and continuum in 13/18 sources. Line luminosities and continuum fluxes were used to estimate molecular gas masses from CO, [C\,\textsc{i}], and dust, and a hierarchical Bayesian cross-calibration of all four tracers yielded consistent per-source $M_{\rm H_2}$ estimates and global conversion factors. Comparison with photodissociation region (PDR) and X-ray dominated region model grids using the $L_{\rm [CII]}/L_{\rm [CI]}$ and $L_{\rm CO(7\text{--}6)}/L_{\rm TIR}$ ratios suggests gas densities of $n > 10^4$~cm$^{-3}$ and radiation fields of $G_0 \sim 10^3$–$10^4$ for the subset of sources consistent with PDR solutions, while many quasars fall outside the model parameter space. Additional diagnostics based on the $L'_{\rm CO(7-6)}/L'_{\rm [CI](2-1)}$ ratio indicate that a large fraction of the molecular gas resides in a warm and highly excited phase. Together these results suggest that classical PDR heating alone cannot explain the observed line ratios and that additional volumetric processes such as X-ray irradiation, turbulence and shocks, or enhanced cosmic-ray heating likely influence the excitation of the cold ISM. These results demonstrate the power of multi-line diagnostics in revealing the excitation and structure of the cold ISM in early quasar host galaxies and highlight the need for a joint analysis of CO, [C\,\textsc{i}], [C\,\textsc{ii}], and dust emission to fully characterize star formation and heating driven by active galactic nuclei at cosmic dawn.
}


\maketitle
\nolinenumbers
\section{Introduction}

The discovery of luminous quasars at redshifts $z \gtrsim 6$ has opened a unique observational window into the evolution of supermassive black holes (SMBHs) and their host galaxies during the first billion years of cosmic history \citep{Fan2000, Fan2003, Fan2006}. Up to now, more than 500 quasars have been spectroscopically confirmed at $z > 6$ \citep[e.g.,][]{Jiang2015, Banados2018, Wang2021, Yang2020, Fan2023}, including several beyond $z \sim 7.5$ \citep{Banados2018, Wang2021, Yang2020}. These systems typically harbor SMBHs with masses $\gtrsim10^9\,M_\odot$, growing at accretion rates exceeding $\dot{M}_{\rm BH} \gtrsim 10\,M_\odot$\,yr$^{-1}$ \citep{Schindler2020, Yang2021, Farina2022}. Their host galaxies are sites of vigorous star formation, with star formation rates (SFRs) typically ranging from 100 to 1000~$M_\odot$\,yr$^{-1}$, as revealed by far-infrared (FIR) and submillimeter observations \citep{Bertoldi2003, Wang2013, Decarli2018, Venemans2020, Wang2024, Bouwens2025}. The host galaxies themselves are also massive systems in most cases, with stellar or dynamical masses on the order of $10^{10}$--$10^{11}\,M_\odot$ \citep[e.g.,][]{Venemans2016, Neeleman2021, Fei2025}. These dual processes place immense demand on the molecular gas reservoirs ($M_{\rm gas} \sim 10^{10}$--$10^{11}\,M_\odot$), requiring efficient replenishment via cold gas accretion from the circumgalactic medium or frequent mergers with gas-rich companions \citep{Trakhtenbrot2017, Farina2019, Decarli2017, Decarli2024, Ding2025}. These findings indicate that both the black holes and their host galaxies underwent rapid growth during the first billion years of cosmic history.

To understand the fueling of star formation and SMBH growth, a quantitative characterization of the cold interstellar medium (ISM)—defined as gas with $T \lesssim 100$\,K—is essential. This cold ISM represents the immediate reservoir available for galaxy-scale activity, and its properties (mass, temperature, density, and ionization state) govern the efficiency and mode of baryonic conversion into stars and black hole accretion.

The Atacama Large Millimeter/submillimeter Array (ALMA) has revolutionized studies of the cold ISM in high-$z$ quasars. At the time of the comprehensive review by \citet{Carilli2013}, only two quasars at $z > 6$ (J1148+5251 and J1120+0641) had been detected in the [C\,\textsc{ii}]~158\,$\mu$m fine-structure line. Today, the number of [C\,\textsc{ii}]-detected quasars at $z > 6$ has grown to almost 200 \citep{Decarli2018, Venemans2020, Wang2024, Bouwens2025}, enabling the first statistical studies of the ISM at the end of the reionization era. [C\,\textsc{ii}] emission provides accurate redshifts, constraints on gas dynamics, and indirect SFR estimates. The observed diversity in [C\,\textsc{ii}] luminosities, spatial extents, line profiles, and associated dust continuum points to a wide range of evolutionary stages and ISM conditions in these young galaxies \citep{Decarli2018, Venemans2019, Neeleman2021, Shao2022}.

However, [C\,\textsc{ii}] alone cannot fully characterize the cold molecular gas. [C\,\textsc{ii}] emission mostly arises from photon-dominated regions (PDRs) in luminous quasar--starburst systems \citep{Diaz-Santos2017,Pensabene2021}, but it can also have significant contributions from diffuse atomic and ionized gas in more quiescent systems, as well as from CO-poor low-metallicity molecular reservoirs, which introduces an intrinsic uncertainty when using [C\,\textsc{ii}] as a tracer of cold molecular gas.  Neutral carbon fine-structure lines, [C\,\textsc{i}]~($^3P_1 \rightarrow {}^3P_0$ and $^3P_2 \rightarrow {}^3P_1$), provide an independent tracer of cold molecular gas \citep{Papadopoulos2004, 	Jiao2017, Decarli2022, FriasCastillo2025}. 
Their excitation energies ($E/k \approx 24$ and $63$~K) are readily reached in warm quasar host environments. Moreover, their moderate critical densities ($n_{\rm crit} \approx 470$ and $1.2 \times 10^3$~cm$^{-3}$) and typically optically thin nature make [C\,\textsc{i}] a reliable tracer of neutral gas closely associated with the low-$J$ CO-emitting molecular component. Nevertheless, both [C\,\textsc{ii}] (with $n_{\rm crit} \sim 2.8 \times 10^3$~cm$^{-3}$) and [C\,\textsc{i}] are characterized by moderate critical densities, which limits their ability to probe denser ($n_{\rm H_2} \gtrsim 10^4$~cm$^{-3}$) and warmer molecular gas phases. 
Such phases are expected to be prominent in luminous $z \sim 6$ quasars, owing to intense star formation and powerful central active galactic nuclei (AGN) activity. Additional diagnostics from mid-$J$ CO transitions are therefore particularly valuable. 
These lines trace highly excited molecular gas that can be heated not only by PDRs but also by X-ray–dominated regions (XDRs), cosmic rays, and shocks \citep{Papadopoulos2004,Meijerink2007,Vallini2018,Riechers2009a,Riechers2020,Decarli2022}. 
Among them, CO(7--6) lies close to the peak of the CO spectral line energy distribution in high-redshift quasars \citep[e.g.,][]{Carilli2013}, making it a sensitive tracer of gas excitation and a useful indicator for estimating CO(1–0)-based molecular gas masses, albeit with uncertainties due to excitation variations.

These various tracers—[C\,\textsc{ii}], CO, [C\,\textsc{i}], and the FIR dust continuum—not only probe different phases and excitation conditions of the cold ISM but also provide complementary approaches to estimate the molecular gas mass in high-$z$ galaxies. Leveraging these diagnostics enables a more complete and robust characterization of the cold gas reservoir that fuels both star formation and black hole accretion. Several techniques have been developed to estimate the total molecular gas mass in high-$z$ galaxies. These include: (i) the FIR dust continuum, calibrated via assumptions about dust temperature and gas-to-dust ratio; (ii) low-$J$ CO luminosities, assuming a CO(1–0) conversion factor $\alpha_{\rm CO}$; (iii) [C\,\textsc{i}] line luminosities, using an abundance ratio $X_{\rm CI}$; and (iv) [C\,\textsc{ii}], adopting a conversion factor $\alpha_{\rm [CII]}$ that links $L_{\rm [CII]}$ to molecular gas mass, though its use is complicated by its multiphase origin \citep{Carilli2013, Walter2011, Zanella2018, Decarli2020, Dunne2021, Neeleman2021, Sommovigo2021, Gururajan2023}. Each method carries systematic uncertainties, and cross–calibrating between them is critical. For example, \citet{Dunne2021} employed a joint modeling approach using CO, [C\,\textsc{i}], and dust to constrain gas masses in star–forming galaxies; \citet{Neeleman2021} used spatially resolved [C\,\textsc{ii}] kinematics to derive dynamical masses of $z\gtrsim6$ quasars and benchmark gas–mass estimators; \citet{Sommovigo2021, Sommovigo2022} used [C\,\textsc{ii}] as a dust–temperature diagnostic to refine gas–mass estimates; and \citet{Decarli2022} combined [C\,\textsc{ii}], [C\,\textsc{i}], CO(7–6), and dust to compare the relative strengths and assumptions of each method in $z\sim6$ quasar hosts; \citet{Gururajan2023} cross–calibrated CO, [C\,\textsc{i}], [C\,\textsc{ii}], and dust tracers in a sample of lensed dusty star-forming galaxies (DSFGs) to quantify systematic offsets; and \citet{DEugenio2023} evaluated [C\,\textsc{ii}] as a gas–mass tracer in high–$z$ quiescent systems, clarifying the regimes where it is most effective.

Line ratios among CO, [C\,\textsc{i}], [C\,\textsc{ii}], and the FIR dust continuum also provide constraints on the physical state of the ISM, probing molecular gas excitation and density as well as the relative importance of different heating mechanisms. For instance, combinations of mid- and high-$J$ CO lines with the FIR continuum primarily trace gas excitation, density, and the intensity of the heating field \citep{Meijerink2007,Carilli2013,Pensabene2021,Xu2024, Tadaki2025}, and ratios between the two [C\,\textsc{i}] fine-structure lines primarily probe the excitation temperature and level populations of neutral carbon \citep{Weiss2005a, Papadopoulos2022}, while the [C\,\textsc{ii}]-to-FIR ratio has long been used as an empirical indicator of star-formation compactness and radiation-field intensity, including the so-called [C\,\textsc{ii}] deficit at high infrared surface brightness \citep{Diaz-Santos2017,Herrera-Camus2018}. When considered jointly, these observables can be compared with theoretical models by invoking far-ultraviolet (FUV)-heated PDRs, XDRs, and additional heating mechanisms such as cosmic rays and mechanical heating \citep{Meijerink2007, Carilli2013, Herrera-Camus2018, Decarli2022, Kaasinen2024}. But in unresolved galaxy-integrated measurements, they remain intrinsically degenerate and should be interpreted as qualitative consistency checks rather than definitive discriminants of the dominant heating mechanism.

In this work, we present a systematic ALMA survey targeting CO(7–6) and [C\,\textsc{i}] emission in a sample of 18 quasars at $z \sim 6$. These sources were selected from the broader population of [C,\textsc{ii}]-detected quasars and span roughly one order of magnitude in infrared luminosity, more than five orders of magnitude in UV luminosity, and about one order of magnitude in the ratio of [C\,\textsc{ii}] to total infrared (TIR) luminosity (see Table~\ref{tab:cii_sample}). Our program builds on the NOEMA pilot study by \citet{Decarli2022} by increasing both the sample size and sensitivity, enabling statistical analysis of ISM conditions and molecular gas reservoirs. Our main goals are to (i) constrain molecular gas masses using multiple tracers, (ii) characterize gas excitation via line ratios, and (iii) investigate the coupling between AGN activity and host galaxy ISM properties in the early Universe.

This paper is structured as follows: Section~\ref{sec:obs} provides a summary of the observations and data reduction; Section~\ref{sec:results} presents line detections and widths; Section~\ref{sec:discussion} presents our estimates of line and infrared luminosities, $M_{\mathrm{H}_2}$, based on various methods along with our constraints on the physical properties of the ISM; and Section~\ref{sec:conclusions} provides a summary and our outlook. Throughout this work, we adopt a \(\Lambda\) cold dark matter cosmology with $H_0 = 70\,\mathrm{km\,s^{-1}\,Mpc^{-1}}$, $\Omega_{\rm m} = 0.3$, and $\Omega_\Lambda = 0.7$.

\begin{figure}[t]
  \centering
  \includegraphics[width=\linewidth]{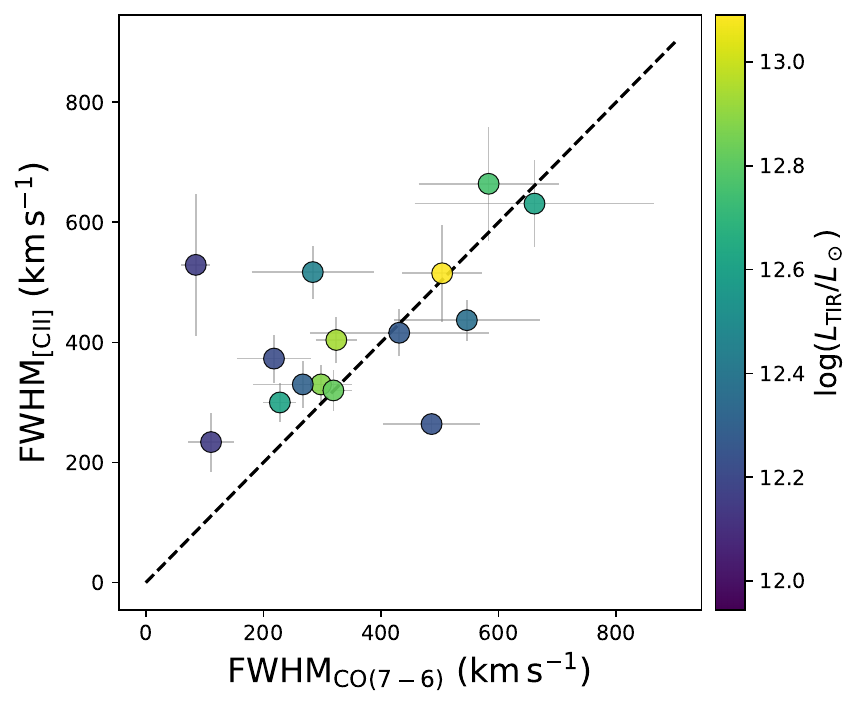}
  \caption{Comparison of velocity widths measured from the two lines.
  The y-axis shows $\mathrm{FWHM}_{\rm [CII]}$ and the x-axis $\mathrm{FWHM}_{\rm CO(7\text{--}6)}$.
  Points are color coded by total infrared luminosity, $\log(L_{\rm TIR}/L_\odot)$ (color bar at right).
  Error bars indicate $1\sigma$ uncertainties. The dashed black line marks the one-to-one relation,
  $\mathrm{FWHM}_{\rm [CII]}=\mathrm{FWHM}_{\rm CO(7\text{--}6)}$.}
  \label{fig:fwhm_colorby_lir}
\end{figure}

\section{Observations and data reduction} \label{sec:obs}
The observations were carried out under ALMA project 2019.1.00147.S (PI: R.~Decarli) between October 2019 and March 2020 using 43--48 ALMA 12~m antennas in the C43-4 configuration. The sources were observed using ALMA Band~3, covering the redshifted frequencies of the [C\,\textsc{i}] ($^3P_1 \rightarrow {}^3P_0$) and CO($J=7\rightarrow6$) lines as well as the 3~mm dust continuum. This yielded an angular resolution of $\sim$1.0$''$--2.0$''$, corresponding to physical scales of $\sim$6--12~kpc at $z\sim6$.

The correlator was configured with four spectral windows (SPWs), each with a bandwidth of 1.875~GHz and a spectral resolution of $\sim$2~MHz ($\sim$5~km\,s$^{-1}$). One SPW (0) was centered on the redshifted CO(7--6) and [C\,\textsc{i}] lines, while the other three (SPWs 1, 2, and 3) were placed in line-free regions to estimate the continuum, except for J1120+0641 ($z=7.0842$), where the CO(6--5) line falls within SPW~3 due to its high redshift, enabling a simultaneous observation of CO(6--5), CO(7--6), and [C\,\textsc{i}].

Each target was observed for a total on-source integration time of $\sim$30--60 minutes, reaching typical sensitivities of $\sim$8--16~$\mu$Jy\,beam$^{-1}$ for the continuum and $\sim$0.18--0.27~mJy\,beam$^{-1}$ per 50~km\,s$^{-1}$ channel in the line cubes. The observational details are summarized in Table~\ref{tab:obs}.

The data were calibrated using the standard ALMA pipeline in version 5.6.1-8 of the Common Astronomy Software Applications (CASA; \citealt{McMullin2007}) package. The calibration included standard procedures for bandpass, flux density, and amplitude/phase calibration.

Imaging was performed using the \texttt{tclean} task in CASA, adopting natural weighting to maximize sensitivity. The data cubes were resampled in velocity space using a linear interpolator, with a default channel width of 50~km\,s$^{-1}$. For source J1104+2134, a broader channel width of 80~km\,s$^{-1}$ was adopted to further improve sensitivity. The cell size was automatically computed using the \texttt{au.pickCellSize} utility script, with the desired number of pixels per synthesized beam set to 5. And we set an image size of 512$\times$512 pixels. The imaging process includes the following steps.
\begin{enumerate}
    \item Continuum imaging: A pure-continuum map was created by combining the line-free channels in SPWs~0, along with the full available bandwidth from SPWs~1, 2, and 3. The line-free channels in SPWs~0 were determined by computing the expected line center frequencies of the [C\,\textsc{i}] and CO(7--6) transitions using the systemic redshift from [C\,\textsc{ii}], and masking out a frequency range of $\pm$0.75\,FWHM$_{\rm [CII]}$ around the line peaks. Imaging was performed using the \texttt{tclean} task in CASA with the  \texttt{auto-multithresh} parameters. 
    
    \item Line cube imaging: The pure-line visibilities were obtained by subtracting the continuum emission from the calibrated data using \texttt{uvcontsub}, with a first-order polynomial fit (\texttt{fitorder=1}) to the line-free channels to allow for a small continuum slope across the spectral window. We then imaged the line data cubes for SPWs~0, 1, 2, and 3 using the same \texttt{auto-multithresh} cleaning setup, except with \texttt{negativethreshold} increased to 15.0 to ensure the masking algorithm does not exclude potential faint emission near the noise level.

    \item Moment map creation and spectrum extraction: The moment-0 (integrated intensity) maps were first generated by collapsing a velocity range of $\pm$0.75\,FWHM$_{\rm [CII]}$ centered on the expected frequency of [C\,\textsc{i}] and CO(7--6), estimated from the [C\,\textsc{ii}] redshift. These maps were refined iteratively based on the best-fit centroid velocity and line width obtained from 1D spectral fitting (see below). Spectra were extracted from the single pixel corresponding to the peak of the line emission in the collapsed maps, as the sources are spatially unresolved at our angular resolution. We fit the extracted spectra with a single-Gaussian to each detected line, depending on whether both [C\,\textsc{i}] and CO(7--6) were detected or blended. If no significant detection was found in the maps, the 1D spectra were instead extracted at the optical source position. The centroid and width from the fitted spectrum were then used to update the integration window for the moment maps, and the process was repeated until convergence.
\end{enumerate}

\begin{figure}[t]
  \centering
  \includegraphics[width=\linewidth]{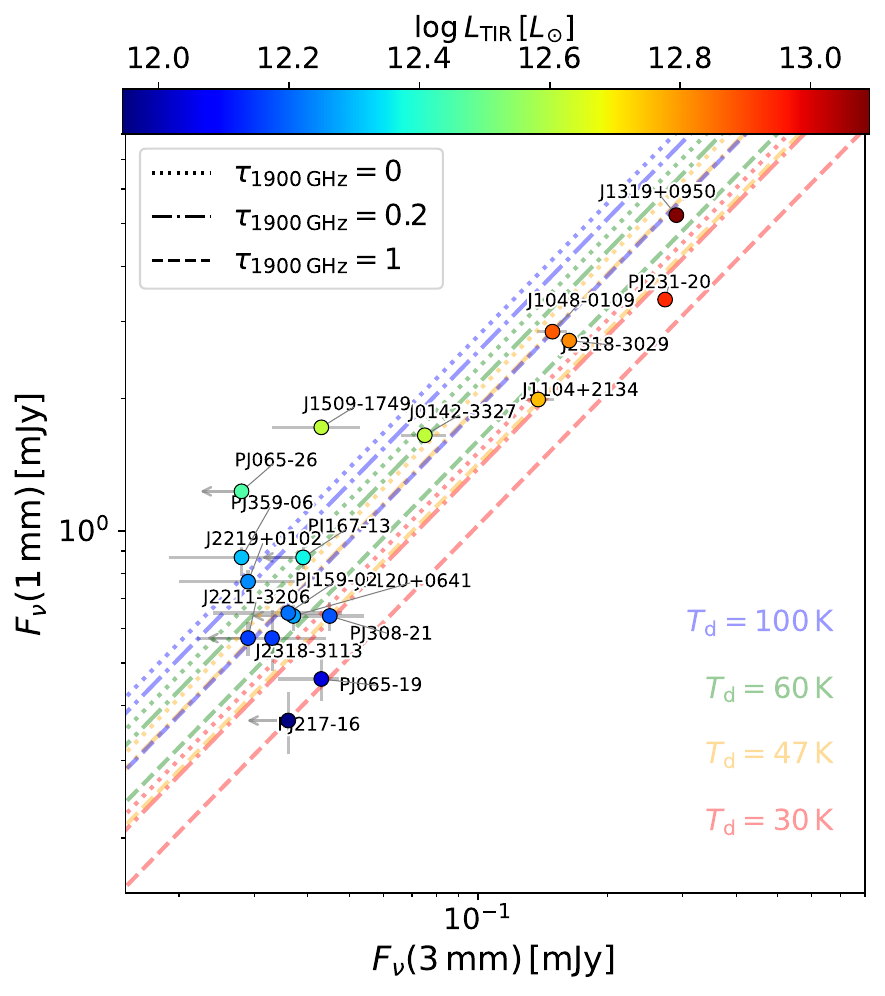}
  \caption{Observed continuum flux densities at 3\,mm and 1\,mm for the $z\!\sim\!6$ quasar sample. Points are color mapped by $\log L_{\rm TIR}$; error bars are shown when available, and gray arrows denote limits. Overplotted are modified blackbody tracks that span dust temperatures $T_{\rm d}=\{30,47,60,100\}\,$K (colors) and an opacity index $\tau_{1900\,{\rm GHz}}=\{0.2,1,5\}$ (linestyles); the CMB at $z=6$ is included. The locus of the data is broadly consistent with $T_{\rm d}\approx47$--$60$\,K and $\tau_{1900\,{\rm GHz}}\approx0.2$--1. In the following, we adopt the $T_{\rm d}=47$\,K, $\tau_{1900\,{\rm GHz}}=0.2$ template when deriving IR luminosities and dust masses.}
  \label{fig:mm-cont}
\end{figure}

\section{Results} \label{sec:results}

As described in the previous section, we first searched the collapsed line maps for sources and extracted the spectra at the position of the emission peaks. For objects where no clear detection associated with the quasar host galaxy was found, we extracted the spectra at the nominal target positions based on optical/NIR coordinates.

We fitted the spectra using least-squares minimization, adopting either a single or double Gaussian profile depending on whether CO(7--6), [C\,\textsc{i}], or both emission was detected. The resulting line parameters (peak frequency, line width, and integrated flux) are reported in Table~\ref{tab:line}. The spectra extracted over SPW~0 for 18 quasars in our sample are provided in the supplementary Figure~\ref{fig:group1}–\ref{fig:group6}.

Given the relatively low signal-to-noise ratio (S/N) in many cases, we consider a line to be detected only if the moment-0 map shows at least a $2\sigma$ detection. If the moment-0 map shows a $>3\sigma$ detection within one ALMA synthesized beam centered at the optical pointing position, we classify it as a significant detection; if the signal is at the 2–3$\sigma$ level within this region, we regard it as a tentative detection. If a $>3\sigma$ feature appears outside one beam from the pointing center, we label it as a potential companion, which is not further discussed in this paper. 

Based on these criteria, CO(7--6) is not detected in three out of the 18 targeted quasars: PJ308--21, PJ217--16, and PJ065--19. Only tentative detections are found for J2318--3113 and PJ065--26, while the remaining 13 sources show significant detections. For [C\,\textsc{i}], only six sources show detections: J1048$-$0109, PJ231--20, J0142$-$3327, PJ359--06, J2318$-$3029, and J1319$+$0950, among which PJ359--06 is only tentatively detected. The depth of the observations for these undetected sources is not significantly different from the rest of the sample. For all non-detections, we estimate a $3\sigma$ upper limit by assuming a Gaussian line profile with a full width at half maximum (FWHM) equal to the [C\,\textsc{ii}] line width reported in the literature.

Figure~\ref{fig:fwhm_colorby_lir} compares the distribution of CO(7--6) line widths (expressed as the FWHM of the Gaussian fits) with the [C\,\textsc{ii}] line widths compiled from previous studies. Due to the limited spectral resolution, the uncertainties on CO(7--6) line widths are relatively large, but the majority of sources show consistent widths between CO(7--6) and [C\,\textsc{ii}]. An exception is J2211$-$3206, which displays a very narrow CO(7--6) line width of $85 \pm 25$\,km\,s$^{-1}$, significantly smaller than the [C\,\textsc{ii}] width of $529 \pm 118$\,km\,s$^{-1}$. Given the low S/N  in the spectrum, this detection may be spurious. 

For the continuum emission, we adopt the same detection criteria as for the line emission. Under this criterion, five quasars show non-detections: J1120$+$0641, PJ217--16, PJ167--13, J2211$-$3206, and PJ065--26. Two sources, J2318--3113 and PJ359--06, show tentative detections, while the remaining eleven quasars exhibit significant detections. If we adopt a more conservative threshold of $5\sigma$, an additional five sources fall below this limit: PJ159--02, J2318--3113, J2219$+$0102, PJ359--06, and J1509--1749.

\begin{table*}
\centering
\caption{Observed properties of the CO(7--6) and [C\,\textsc{i}] emission lines.\label{tab:line}}
\begin{tabular}{lc|cc|cc|cc}
\hline
Target & $F_{\rm cont}$(3mm)
& \multicolumn{2}{c|}{Redshift} 
& \multicolumn{2}{c|}{FWHM [km\,s$^{-1}$]} 
& \multicolumn{2}{c}{$F_{\rm line}$ [Jy\,km\,s$^{-1}$]} \\
&[mJy] &$z_{\rm CO}$ & $z_{\rm [CI]}$ 
& CO & [C\,\textsc{i}] 
& CO & [C\,\textsc{i}] \\
\hline
J1120+0641 & $< 0.037$ &$7.0875 \pm 0.0017$ & $\cdots$ & $431 \pm 152$ & $\cdots$ & $0.183 \pm 0.056$ & $< 0.091$ \\
J1104+2134 & $0.138 \pm 0.012$ &$6.7676 \pm 0.0013$ & $\cdots$ & $583 \pm 119$ & $\cdots$ & $0.349 \pm 0.062$ & $< 0.086$ \\
J1048-0109 & $0.149 \pm 0.012$ &$6.6757 \pm 0.0004$ & $6.6760 \pm 0.0015$ & $298 \pm 37$ & $319 \pm 133$ & $0.438 \pm 0.047$ & $0.133 \pm 0.048$ \\
PJ231-20 & $0.273 \pm 0.010$ &$6.5867 \pm 0.0004$ & $6.5871 \pm 0.0009$ & $324 \pm 35$ & $257 \pm 81$ & $0.46 \pm 0.043$ & $0.134 \pm 0.036$ \\
PJ167-13 & $< 0.039$ &$6.5181 \pm 0.0013$ & $\cdots$ & $546 \pm 124$ & $\cdots$ & $0.297 \pm 0.058$ & $< 0.113$ \\
J2318-3113 & $0.033 \pm 0.011^t$ &$6.4453 \pm 0.0004$ & $\cdots$ & $111 \pm 39$ & $\cdots$ & $0.068 \pm 0.021^t$ & $< 0.076$ \\
PJ159-02 & $0.036 \pm 0.012$ &$6.3852 \pm 0.0007$ & $\cdots$ & $218 \pm 63$ & $\cdots$ & $0.154 \pm 0.038$ & $< 0.102$ \\
J2211-3206 & $< 0.029$ &$6.3375 \pm 0.0003$ & $\cdots$ & $85 \pm 25$ & $\cdots$ & $0.066 \pm 0.018$ & $< 0.102$ \\
J0142-3327 & $0.075 \pm 0.009$ &$6.3372 \pm 0.0003$ & $6.3375 \pm 0.0012$ & $228 \pm 28$ & $376 \pm 119$ & $0.282 \pm 0.03$ & $0.141 \pm 0.039$ \\
PJ308-21 & $0.045 \pm 0.009$ &$\cdots$ & $\cdots$ & $\cdots$ & $\cdots$ & $< 0.094$ & $< 0.101$ \\
PJ065-26 & $< 0.028$ &$6.1886 \pm 0.0011$ & $\cdots$ & $284 \pm 104$ & $\cdots$ & $0.123 \pm 0.039^t$ & $< 0.109$ \\
PJ359-06 & $0.028 \pm 0.009^t$ &$6.1714 \pm 0.0009$ & $6.1715 \pm 0.0016$ & $267 \pm 84$ & $226 \pm 154$ & $0.125 \pm 0.034$ & $0.053 \pm 0.031^t$ \\
PJ217-16 & $< 0.036$ &$\cdots$ & $\cdots$ & $\cdots$ & $\cdots$ & $< 0.118$ & $< 0.127$ \\
J2219+0102 & $0.029 \pm 0.009$ &$6.1484 \pm 0.0008$ & $\cdots$ & $486 \pm 83$ & $\cdots$ & $0.313 \pm 0.046$ & $< 0.079$ \\
J2318-3029 & $0.163 \pm 0.008$ &$6.1466 \pm 0.0003$ & $6.1447 \pm 0.0008$ & $319 \pm 32$ & $270 \pm 78$ & $0.389 \pm 0.034$ & $0.14 \pm 0.035$ \\
J1319+0950 & $0.290 \pm 0.012$ &$6.1338 \pm 0.0007$ & $6.1329 \pm 0.0012$ & $504 \pm 68$ & $413 \pm 122$ & $0.595 \pm 0.069$ & $0.245 \pm 0.063$ \\
PJ065-19 & $0.043 \pm 0.009$ &$\cdots$ & $\cdots$ & $\cdots$ & $\cdots$ & $< 0.091$ & $< 0.095$ \\
J1509-1749 & $0.043 \pm 0.010$ &$6.1227 \pm 0.0021$ & $\cdots$ & $661 \pm 204$ & $\cdots$ & $0.244 \pm 0.065$ & $< 0.136$ \\
\hline
\end{tabular}
\tablefoot{$t$ indicates tentative detections at the 2–3$\sigma$ significance level.}
\end{table*}

\begin{figure}
    \centering
    \includegraphics[width=\linewidth]{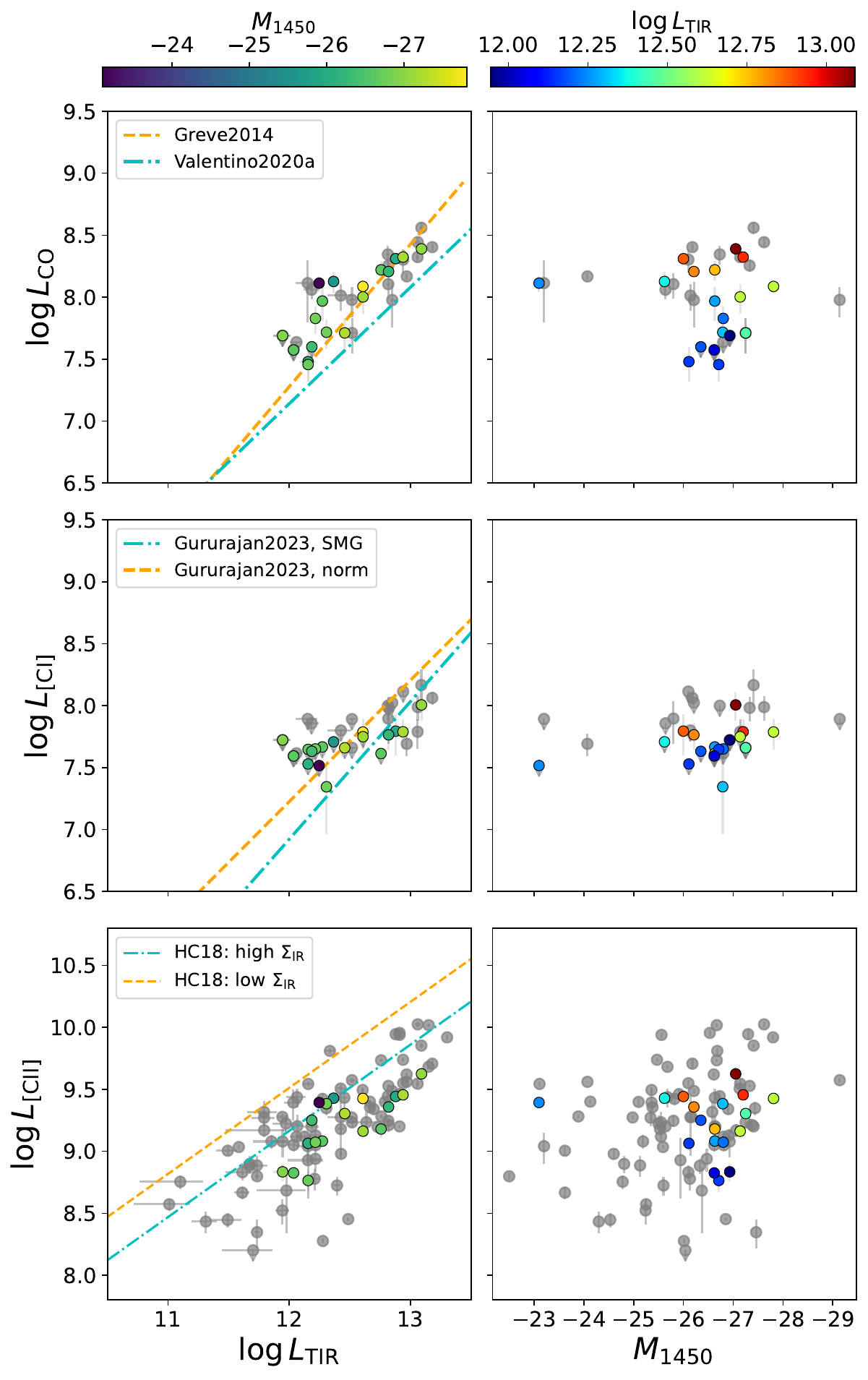}
    \caption{
    Comparison between FIR and UV luminosities with molecular and atomic line luminosities for the quasar sample.
    Each row corresponds to a different emission line: from top to bottom, CO(7–6), [C\,\textsc{i}], and [C\,\textsc{ii}].
    The left column shows the correlation with total infrared luminosity ($\log L_{\rm TIR}$), while the right column shows the correlation with rest-frame UV magnitude ($M_{1450}$).
    Data points are color coded by the complementary variable (i.e., colored by $M_{1450}$ in the left column, and by $\log L_{\rm TIR}$ in the right column).
    Error bars are shown where available, and upper limits are indicated with gray arrows. Gray circles denote all currently published measurements at $z>5.7$ compiled from the literature. For CO(7–6), we compare with the best-fit relation from the local (U)LIRG sample of \citet{Greve2014} (orange dashed line) and from the main-sequence galaxies and starbursts at $z \sim 1.3$ of \citet{Valentino2020a}, which also includes mapped nearby objects from \citet{Liu2015} (cyan dash-dotted line). For [C\,\textsc{i}], we compare with the fits from all SMGs at $z=2$–4 of \citet{Gururajan2023} (cyan dash-dotted line) and the fit for main-sequence galaxies at $z \sim 1$ (orange dashed line). For [C\,\textsc{ii}], we compare with the two average values of the [C\,\textsc{ii}]/IR luminosity ratio for low–$\Sigma_{\rm IR}$ (orange dashed line) and high–$\Sigma_{\rm IR}$ galaxies (cyan dash-dotted line) from \citet{Herrera-Camus2018}.
    }
    \label{fig:luminosity_comparison}
\end{figure}

\section{Discussion}  \label{sec:discussion}
\subsection{Continuum and line luminosities} \label{sec:luminosity}

In order to assess the physical conditions of the cold ISM and estimate gas properties, we derived the luminosities of the dust continuum, CO(7--6), and [C\,\textsc{i}] emission lines for each source in our sample. To compute the total infrared luminosity and constrain the dust properties, we fit the observed continuum flux densities using a modified blackbody function that accounts for both the cosmic microwave background (CMB) and the optical depth of dust emission. The observed flux density at frequency $\nu_{\rm obs}$ is expressed as

\begin{equation}
S_{\nu, \mathrm{obs}} = \frac{\Omega}{(1+z)^3} \left[ B_\nu(T_{\rm d}) - B_\nu(T_{\mathrm{CMB}}) \right] \left[1 - \exp(-\tau_\nu) \right], \label{eq:snuobs}
\end{equation}

where $\Omega$ is the solid angle subtended by the source, $T_{\rm d}$ is the dust temperature, and $T_{\mathrm{CMB}} = 2.725\,(1+z)$\,K is the CMB temperature at redshift $z$. $B_\nu(T)$ is the Planck function at rest-frame frequency $\nu$ and temperature $T$. The optical depth $\tau_\nu$ is given by

\begin{equation}
\tau_\nu = \kappa_{\rm d}(\nu)\,\Sigma_{\rm d} \approx \kappa_{\rm d}(\nu) \left( \frac{M_{\rm d}}{\Omega D_A^2} \right) = \kappa_{\rm d}(\nu) \left( \frac{N_{\mathrm{H}_2}\,m_{\mathrm{H}_2}}{\delta_{\mathrm{GDR}}} \right), \label{eq:taunu}
\end{equation}

where $\kappa_{\rm d}(\nu)$ is the dust mass absorption coefficient, $M_d$ is the dust mass, $D_A$ is the angular diameter distance, and $\delta_{\mathrm{GDR}}$ is the gas-to-dust mass ratio. For optically thin emission ($\tau_\nu \ll 1$), Eq.~\ref{eq:snuobs} simplifies to

\begin{equation}
S_{\nu,\mathrm{obs}} \approx \frac{(1+z)}{D_L^2} \, \kappa_{\rm d}(\nu) \, M_{\rm d} \left[ B_\nu(T_{\rm d}) - B_\nu(T_{\mathrm{CMB}}) \right]. \label{eq:optthin}
\end{equation}

We followed \citet{Decarli2022} for the dust–opacity law, adopting $\kappa_{\rm d}(\nu)$ obtained by interpolating the tabulated values in \citet[][Table~5]{Draine2003} across the infrared, and extrapolating with a power law $\kappa_{\rm d}\propto\nu^{\beta}$ using $\beta=1.9$ for $\nu\lesssim 350~\mathrm{GHz}$ (see also \citealt{daCunha2021}).

In Fig.~\ref{fig:mm-cont} we compare the 1\,mm continuum flux density (evaluated at the observed frequency of [C\,\textsc{ii}], following the literature references compiled in Table~\ref{tab:cii_sample}) against the observed 3\,mm flux density measured at the frequency of the CO(7--6) transition (from our ALMA observations). We do not attempt to fit for $T_{\rm d}$, $\beta$, or $\tau_{\rm d}$. We merely overplot a plausible graybody model to illustrate that the combination of our 1\,mm and 3\,mm continuum points does not contradict typical dust temperatures inferred for similar systems. Relative to \citet{Decarli2022}, our sample shows a larger dispersion in the 1\,mm/3\,mm color; for J1509$-$1749 and PJ065$-$26 the extreme ratios would require $\tau\simeq 0$ and very high dust temperatures ($T_{\rm d}\gg 100$\,K) to match simple modified–blackbody tracks. As illustrated in Fig.~\ref{fig:ratio-hist}, the median observed continuum ratio is about 18, consistent with the typical value of about 19 reported by
\citet{Decarli2022}. Such colors are reproduced by $(T_{\rm d},\,\tau_{1900\,\mathrm{GHz}})=(60~\mathrm{K},\,1)$ or $(47~\mathrm{K},\,0.2)$. Overall, the continuum ratios are consistent with optically thin dust emission with characteristic temperatures of $T_{\rm d} \sim 37$\,K. Nevertheless, the extreme colors of J1509$-$1749 and PJ065$-$26 deviate from these trends, possibly reflecting unusually warm dust, a flatter emissivity index, or additional emission components at millimeter wavelengths.

Given that many sources are only weakly detected (or undetected) at 3\,mm, and to ensure consistency with \citet{Decarli2022}, we derive $L_{\rm TIR}$ ($8$--$1000\,\mu\mathrm{m}$) from the 1\,mm flux density alone,
adopting $(T_{\rm d},\,\tau_{1900\,\mathrm{GHz}})=(47~\mathrm{K},\,0.2)$. For all comparison $z\sim6$ quasars used later, we refit $L_{\rm TIR}$ uniformly with the same procedure.

We computed the line luminosity in units of brightness temperature and velocity-integrated flux ($L'$; K\,km\,s$^{-1}$\,pc$^2$) using the standard relation \citep{Carilli2013}:

\begin{equation}
L' = 3.25 \times 10^7 \, S_{\rm line} \Delta v \, \nu_{\rm obs}^{-2} \, D_L^2 \, (1+z)^{-3},
\end{equation}

where $S_{\rm line} \Delta v$ is the velocity-intergrated line flux in Jy\,km\,s$^{-1}$, $\nu_{\rm obs}$ is the observed line frequency in gigahertz, and $D_L$ is in megaparsecs. The corresponding line luminosity in solar units is

\begin{equation}
L = 1.04 \times 10^{-3} \, S_{\rm line} \Delta v \, \nu_{\rm rest} \, D_L^2 \, (1+z)^{-1}, \label{eq:Lsun}
\end{equation}

where $\nu_{\rm rest}$ is the rest-frame frequency in gigahertz, and $L$ is in $L_\odot$.

In Figure~\ref{fig:luminosity_comparison} we compare the luminosities of the three lines under study with the total infrared luminosity and with the rest–frame UV magnitude $M_{1450}$.  The gray circles mark a literature compilation of all published $z>5.7$ quasars with CO(7--6) and [C\,\textsc{i}] measurements or limits \citep{Riechers2009a, Venemans2016, Venemans2017, Venemans2017b, Novak2019, Wang2019a, Yang2019, Pensabene2021, Decarli2022, Decarli2023, Fujimoto2022, Feruglio2023, Li2022, Li2024} and with [C\,\textsc{ii}] detections \citep{Walter2009, Venemans2012, Wang2013, Banados2015, Willott2015, Wang2016, Willott2017, Mazzucchelli2017, Venemans2017b, Decarli2018, Feruglio2018, Izumi2018, Decarli2019a, Izumi2019, Neeleman2019, Yang2019, Andika2020, Eilers2020, Yang2020, Venemans2020, Izumi2021, Wang2021, Fujimoto2022, Khusanova2022, Meyer2022, Shao2022, Endsley2023a, Banados2024, Bischetti2024, Izumi2024, Mazzucchelli2024, Wang2024, Zhu2024}; we refer to this comparison set as the \texttt{Allz6quasar} sample. For targets with multiple observations, we use the measurement from the deeper (higher-sensitivity) dataset. Our targets span $\sim$1\,dex in $L_{\rm TIR}$ and a comparable dynamic range in each detected line, with most $M_{1450}$ clustered among luminous quasars ($M_{1450}\lesssim-25$). The \texttt{Allz6quasar} [C\,\textsc{ii}] sample extends over $\sim$2\,dex in both $L_{\rm TIR}$ and $L_{\rm [CII]}$ and covers approximately $-29\lesssim M_{1450}\lesssim-22$.  Because of sensitivity limits, existing CO(7--6) and [C\,\textsc{i}] measurements predominantly probe systems with $L_{\rm TIR}\gtrsim10^{12}\,L_\odot$. Our CO(7--6) and [C\,\textsc{i}] detections span line luminosities and host-galaxy and AGN properties ($L_{\rm TIR}$, $M_{1450}$) comparable to the \texttt{Allz6quasar} sample. The measured CO(7--6) luminosities follow empirical trends calibrated on local ultraluminous infrared galaxies (ULIRGs; e.g., \citealt{Greve2014}) and show systematically higher $L_{\rm CO(7-6)}$ at a given $L_{\rm TIR}$ compared to main-sequence galaxies and starbursts at $z \sim 1.3$ \citep{Valentino2020a}. However, we note that for $z \sim 6$ quasars, including our sample, the steeper $L_{\rm CO(7-6)}$–$L_{\rm TIR}$ relation compared to the ULIRG sample of \citet{Greve2014} indicates enhanced molecular gas excitation, suggesting that excitation up to $J{=}7$ is primarily powered by star formation, though AGN heating may also play a role. Detected [C\,\textsc{i}] luminosities of $(3$–$10)\times10^{7}\,L_\odot$ broadly agree with the [C\,\textsc{i}]–IR relation of all submillimeter galaxies (SMGs) at $z=2$–4 from \citet{Gururajan2023} and with the fit for main-sequence galaxies at $z \sim 1$, although most of our [C\,\textsc{i}] measurements are upper limits and deeper data are required. For comparison, the predicted [C\,\textsc{ii}]/IR ratios in low- and high-$\Sigma_{\rm SFR}$ regimes are $\approx1.6\times10^{-3}$ and $\approx5\times10^{-4}$, respectively \citep{Herrera-Camus2018}, adopting the SFR–IR conversion of \citet{Kennicutt2012}. Most of our $L_{\rm [CII]}$ values lie at or even below the high--$\Sigma_{\rm SFR}$ sequence, implying very compact star formation and strong radiation fields (and possibly enhanced dust opacity or AGN–heated continuum) that raise the IR continuum more rapidly than the [C\,\textsc{ii}] line emission. Overall, all three lines correlate positively with $L_{\rm TIR}$, while no clear trend is seen with $M_{1450}$.

In Figure~\ref{fig:lum_linetoTIR}, we further examine the line-to-TIR ratios, $L_{\rm line}/L_{\rm TIR}$, as functions of $L_{\rm TIR}$ and $M_{1450}$. The decrease of $L_{\rm [CII]}/L_{\rm FIR}$ with increasing $L_{\rm FIR}$—the [C\,\textsc{ii}] deficit—is clearly visible in our sample, with its upper envelope following the high- and low-$\Sigma_{\rm SFR}$ sequences of \citet{Herrera-Camus2018}. A larger fraction of our targets fall below the high--$\Sigma_{\rm SFR}$ sequence, consistent with the luminosity comparison shown in Figure~\ref{fig:luminosity_comparison}. Interestingly, CO(7--6)/TIR and [C\,\textsc{i}]/TIR show similar decreasing trends, whereas main-sequence galaxies and starbursts exhibit flatter or even rising relations. 

In Figure~\ref{fig:lum_ciitoline}, we compare the luminosity ratios between the three lines as functions of $L_{\rm TIR}$ and $M_{1450}$. The $L_{\rm [CII]}/L_{\rm [CI]}$ ratio increases with $L_{\rm TIR}$, while $L_{\rm [CII]}/L_{\rm CO(7-6)}$ remains roughly constant across the sampled range, with no clear dependence on $M_{1450}$. These trends are consistent with the deficit behavior discussed above and further suggest changing excitation and optical-depth conditions in the most IR-luminous quasars. A detailed interpretation and implications for the [C\,\textsc{ii}]-based gas masses are discussed in Section~\ref{sec:gas_mass}.

\begin{table*}
\centering
\caption{Derived line luminosity and luminosity ratios.\label{tab:lum}}
\begin{tabular}{lccccc}
\hline
Target 
& $L_{\rm CO(7-6)}$ 
& $L_{\rm [CI]}$ 
& $\log(L_{\rm [CII]}/L_{\rm [CI]})$ 
& $\log(L_{\rm [CII]}/L_{\rm CO(7-6)})$ 
& $\log(L_{\rm [CI]}/L_{\rm CO(7-6)})$ \\
& [$\rm 10^{8}\, L_\odot$] & [$\rm 10^7\, L_\odot$] & & & \\
\hline
J1120+0641 & $0.93 \pm 0.28 $ & $<4.64 $ & $>1.416$ & $1.114 \pm 0.138$ & $<-0.302$ \\
J1104+2134 & $1.66 \pm 0.29 $ & $<4.11 $ & $>1.567$ & $0.960 \pm 0.103$ & $<-0.607$ \\
J1048-0109 & $2.04 \pm 0.22 $ & $6.23 \pm 2.25 $ & $1.648 \pm 0.157$ & $1.132 \pm 0.048$ & $-0.516 \pm 0.164$ \\
PJ231-20 & $2.10 \pm 0.20 $ & $6.15 \pm 1.65 $ & $1.667 \pm 0.118$ & $1.133 \pm 0.045$ & $-0.534 \pm 0.124$ \\
PJ167-13 & $1.34 \pm 0.26 $ & $<5.11 $ & $>1.721$ & $1.303 \pm 0.086$ & $<-0.418$ \\
J2318-3113 & $0.30 \pm 0.09 $ & $<3.38 $ & $>1.535$ & $1.585 \pm 0.145$ & $<0.050$ \\
PJ159-02 & $0.67 \pm 0.17 $ & $<4.47 $ & $>1.423$ & $1.245 \pm 0.110$ & $<-0.177$ \\
J2211-3206 & $0.29 \pm 0.08 $ & $<4.43 $ & $>1.118$ & $1.309 \pm 0.145$ & $<0.191$ \\
J0142-3327 & $1.22 \pm 0.13 $ & $6.12 \pm 1.69 $ & $1.640 \pm 0.121$ & $1.340 \pm 0.047$ & $-0.300 \pm 0.129$ \\
PJ308-21 & $<0.40 $ & $<4.28 $ & $>1.619$ & $>1.652$ & \phantom{<} $\cdots$ \\
PJ065-26 & $0.51 \pm 0.16 $ & $<4.57 $ & $>1.645$ & $1.594 \pm 0.140$ & $<-0.051$ \\
PJ359-06 & $0.52 \pm 0.14 $ & $2.21 \pm 1.29 $ & $2.039 \pm 0.256$ & $1.668 \pm 0.121$ & $-0.371 \pm 0.280$ \\
PJ217-16 & $<0.49 $ & $<5.28 $ & $>1.112$ & $>1.145$ & \phantom{<} $\cdots$ \\
J2219+0102 & $1.30 \pm 0.19 $ & $<3.28 $ & $>1.878$ & $1.281 \pm 0.068$ & $<-0.596$ \\
J2318-3029 & $1.61 \pm 0.14 $ & $5.81 \pm 1.45 $ & $1.594 \pm 0.111$ & $1.151 \pm 0.044$ & $-0.442 \pm 0.115$ \\
J1319+0950 & $2.45 \pm 0.28 $ & $10.14 \pm 2.61 $ & $1.619 \pm 0.127$ & $1.235 \pm 0.078$ & $-0.384 \pm 0.123$ \\
PJ065-19 & $<0.37 $ & $<3.92 $ & $>1.232$ & $>1.252$ & \phantom{<} $\cdots$ \\
J1509-1749 & $1.00 \pm 0.27 $ & $<5.61 $ & $>1.413$ & $1.161 \pm 0.121$ & $<-0.252$ \\\hline
\end{tabular}
\end{table*}

\begin{figure}[htbp]
    \centering
    \includegraphics[width=\linewidth]{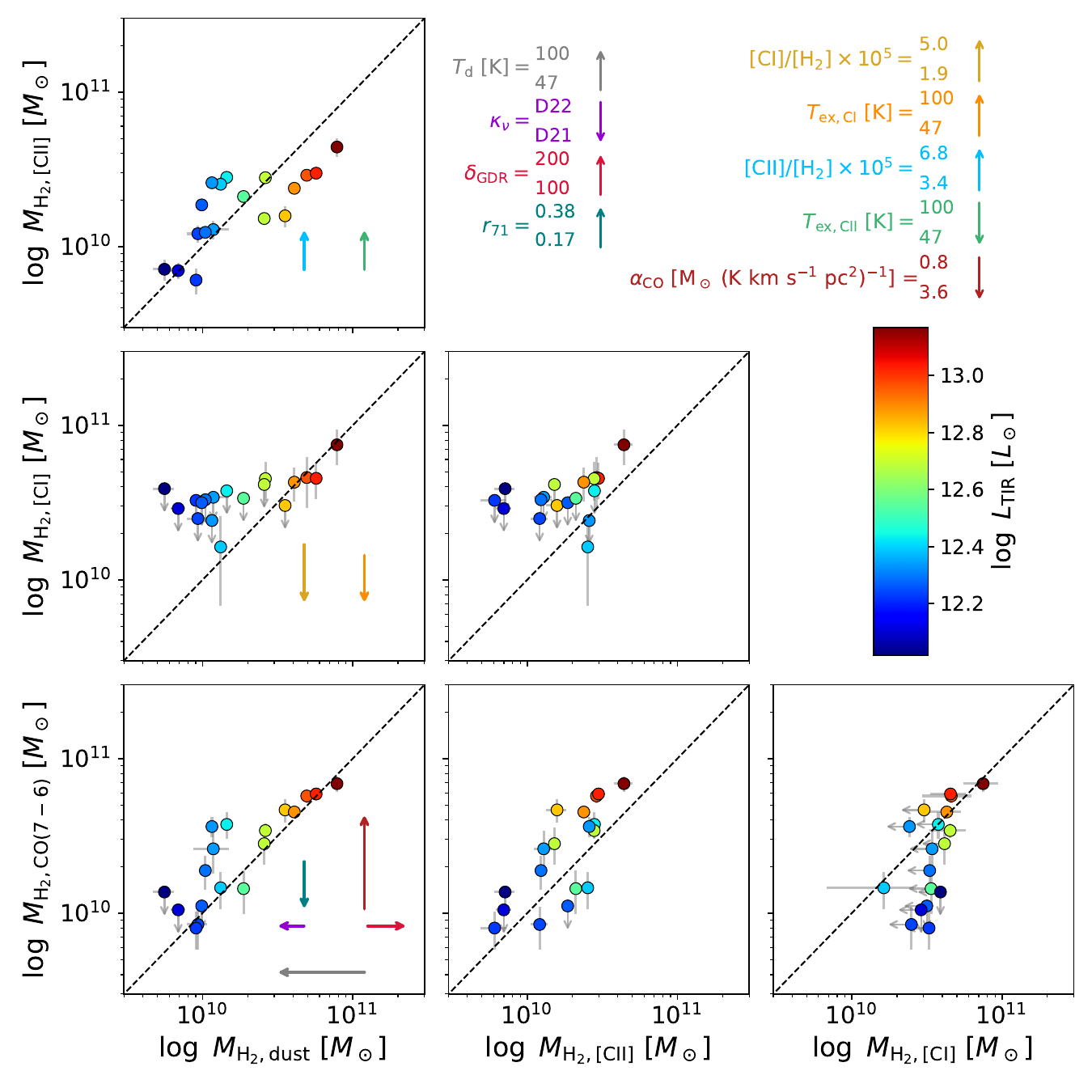}
    \caption{
    Comparison between molecular gas masses estimated from CO(7–6), [C\,\textsc{i}], [C\,\textsc{ii}], and dust continuum emission, following the methods detailed in the main text. The dashed line marks the one-to-one relation. While the overall agreement between tracers is reasonable, systematic deviations appear---particularly for [C\,\textsc{ii}]-based estimates, which tend to yield higher values compared to CO and dust tracers. The arrows illustrate the expected shift in the derived masses when key assumptions (e.g., dust temperature, opacity, gas-to-dust ratio, excitation conditions) are modified from the fiducial choices to values commonly adopted in the literature. Arrow directions correspond to specific parameter changes as summarized in the top-right legend.
    }
    \label{fig:H2_mass_vs_LIR}
\end{figure}

\subsection{Gas mass estimates} \label{sec:gas_mass}

We estimate the molecular gas mass ($M_{\mathrm{H}_2}$) in our sample using four independent tracers: CO(7--6), dust continuum, [C\,\textsc{i}], and [C\,\textsc{ii}] line emission. Each method carries its own assumptions and systematics, allowing us to cross-check results and assess the robustness of our gas mass constraints.

\subsubsection{Single-tracer estimates of molecular gas masses}

We computed the molecular gas mass based on the observed CO(7--6) line luminosity using the following relation:

\begin{equation}
M_{\rm H_2}^{\rm CO(7-6)} = \alpha_{\mathrm{CO}}\, r_{71}^{-1}\, L'_{\mathrm{CO(7\text{--}6)}}, \label{eq:mh2co}
\end{equation}

where we adopted a CO-to-H$_2$ conversion factor $\alpha_{\mathrm{CO}} = 0.8\,M_\odot\,(\mathrm{K\,km\,s^{-1}\,pc^2})^{-1}$, typical for ULIRGs and quasars \citep{Bolatto2013, Carilli2013}. The line ratio $r_{71} = L'_{\mathrm{CO(7\text{--}6)}} / L'_{\mathrm{CO(1\text{--}0)}}$ accounts for CO excitation, and we assume $r_{71} = 0.17$, consistent with high-$z$ quasar templates \citep{Boogaard2020, Decarli2022}. Although CO(7--6) primarily traces warm and dense gas, adopting this excitation ratio allows the line to be used as an empirical proxy for the total molecular gas reservoir.

We derive dust masses $M_{\rm dust}$ by fitting the dust SEDs with the same modified–blackbody model as in Section~\ref{sec:luminosity} ($T_{\rm d}=47$\,K, $\tau_{1900\,\mathrm{GHz}}=0.2$, and the adopted $\kappa_{\rm d}(\nu)$).
Assuming a fixed gas–to–dust ratio ($\delta_{\rm GDR}$) of 100 \citep[see, e.g., ][]{Bolatto2013, Sandstrom2013, Genzel2015, Berta2016, Decarli2022}, the molecular gas mass is
\begin{equation}
M_{\mathrm{H}_2}^{\rm dust}={\delta_{\rm GDR}} \times M_{\rm dust}.
\label{eq:mh2dust}
\end{equation}

This keeps the gas masses consistent with our $L_{\rm TIR}$ estimates. The inferred molecular gas masses are listed in Table \ref{tab:mass}.

The atomic carbon mass is derived from the [C\,\textsc{i}](\,$^3P_2 \rightarrow {}^3P_1$) line luminosity following \citet{Weiss2003, Bothwell2017, Boogaard2020}.

\begin{equation}
M_{\mathrm{CI}} = \frac{4.556 \times 10^{-4}}{Q_{\rm 21}}\, \frac{L'_{\mathrm{[CI](2-1)}}}{\rm K\, km\, s^{-1}\, pc^2}, \label{eq:mci}
\end{equation}

\begin{equation}
Q_{\mathrm{21}} = \frac{5 \exp(-62.5/T_{\rm ex,[CI]})}{1 + 3 \exp(-23.6/T_{\rm ex,[CI]}) + 5 \exp(-62.5/T_{\rm ex,[CI]})}
\end{equation}

where $Q_{\mathrm{21}}$ is the fractional level populations implied by the partition functions at the excitation temperatures $T_{\rm ex,[CI]}$. 
We adopt $T_{\rm ex,[CI]} = 47$\,K, corresponding to the typical dust temperature inferred for high-$z$ quasars \citep{Beelen2006, Leipski2014}, 
implicitly assuming that the [C\,\textsc{i}] levels are in local thermal equilibrium with the dust. In galaxy-integrated observations, [C\,\textsc{i}] emission can trace a substantial fraction of the molecular gas reservoir, as turbulent mixing or cosmic-ray driven chemistry may broaden the classical PDR transition layers and allow [C\,\textsc{i}] to coexist with CO over much of the cloud volume \citep{Papadopoulos2004, Clark2019}. The resulting $M_{\rm H_2}^{\rm [CI]}$ is converted to molecular gas mass assuming a carbon abundance of $X_{\rm CI}\,([\mathrm{C\,\textsc{i}}]/[\mathrm{H}_2]) = (1.9 \pm 0.4) \times 10^{-5}$ \citep{Boogaard2020, Dunne2021}.

To estimate the mass of singly ionized carbon, we followed the formalism in \citet{Weiss2005, Venemans2017b} assuming optically thin emission:

\begin{equation}
M_{\mathrm{C^+}} = \frac{2.92 \times 10^{-4}}{Q_{\rm [CII]}}\, \frac{L'_{\mathrm{[CII]}}}{\rm K\, km\, s^{-1}\, pc^2}, \label{eq:mcii}
\end{equation}

\begin{equation}
Q_{\rm [CII]} = \frac{4\,e^{-91.2/T_{\rm ex,[CII]}}}{2+4\,e^{-91.2/T_{\rm ex,[CII]}}}
\end{equation}

where $Q_{\rm [CII]}$ is the fractional level populations implied by the partition functions at the excitation temperatures $T_{\rm ex,[CII]}$. For PDR gas we adopt $T_{\rm ex,[CII]}=100$\,K, appropriate for [C\,\textsc{ii}]–emitting regions \citep{Meijerink2007,Venemans2017b}. We convert the inferred $M_{\mathrm{C^+}}$ to $M_{\rm H_2}^{\rm [CII]}$  assuming a carbon abundance $X_{\rm C^+}([\mathrm{C^+}]/[\mathrm{H}_2])=3.4\times10^{-5}$. This value is $\sim1.8$ times higher than the $X_{\rm CI}$ abundance adopted by \citet{Boogaard2020}, but remains consistent with other measurements in the literature (e.g., \citealt{Weiss2005}).

Figure~\ref{fig:H2_mass_vs_LIR} compares the molecular--gas masses derived from four tracers (CO(7--6), [C\,\textsc{i}], [C\,\textsc{ii}], and dust continuum) in our sample. Unlike the analysis of \citet{Decarli2022}, where the [C\,\textsc{ii}]--based masses exceed the others by $\simeq0.5$\,dex, our [$\mathrm{C\,\textsc{ii}}$] estimates agree with CO, [C\,\textsc{i}], and dust within the scatter. The key difference is the adopted conversion: \citet{Decarli2022} used a fixed $\alpha_{[\mathrm{CII}]}=30~M_\odot\,L_\odot^{-1}$ (effectively corresponding to $T_{\rm ex,[CII]}=47$\,K and $X_{\rm C^+}=3.4\times10^{-5}$), whereas we assume $T_{\rm ex,[CII]}=100$\,K for PDR gas, which lowers the implied $\alpha_{[\mathrm{CII}]}$ by $\approx0.5$\,dex and reconciles the tracers.

Sensitivity to operative assumptions. The relative alignment (or tension) among the four $M_{\rm H_2}$ estimators depends on a handful of parameters that are not directly measured. For the CO route, the mass scales as $M_{\rm H_2}^{\rm CO(7-6)}\!\propto\!\alpha_{\rm CO}\,r_{71}^{-1}$. Using the updated $\alpha_{\rm CO}=1.7$ for ULIRGs reported by \citet{Arroyave2023} would increase our CO–based masses by roughly a factor of two, reducing the consistency with the other tracers. A still larger value of $\alpha_{\rm CO}$, characteristic of main–sequence disks (e.g., $\alpha_{\rm CO}\!\approx\!3.6~M_\odot\,({\rm K\,km\,s^{-1}\,pc^2})^{-1}$; \citealt{Daddi2010}), would raise the CO–based masses by a factor of about four relative to a ULIRG–like value, further degrading the agreement. Conversely, assuming a higher excitation (e.g., $r_{71}\!\approx\!0.38$ instead of $0.17$; \citealt{Carilli2013}) would lower the CO–based masses by a factor of about 2--2.5, again driving them away from the other tracers.

For [C\,\textsc{i}], the conversion depends on the level population and the carbon abundance. In the absence of direct constraints on the excitation conditions, we adopt the LTE approximation $T_{\rm dust} \sim T_{\rm kin} \sim T_{\rm ex}$. We emphasize that this equality is not physically required and is introduced solely due to the lack of independent measurements of $T_{\rm dust}$, $T_{\rm kin}$, and $T_{\rm ex}$, given that only a single [C\,\textsc{i}] transition is available. Under this assumption, a higher $T_{\rm ex}$ (corresponding to a larger partition function) or a higher carbon abundance $X_{\rm CI}$ would decrease the inferred molecular gas mass $M_{\rm H_2,[CI]}$. For example, adopting $X_{\rm CI} \sim 5 \times 10^{-5}$, as in \citet{Weiss2005}, would result in lower [C\,\textsc{i}]-based gas masses compared to our fiducial choice \citep{Boogaard2020}. We also note that assuming LTE likely overestimates the excitation conditions, as the $Q_{21}$ factor is expected to be lower under sub-thermal excitation of the [C\,\textsc{i}](2--1) transition \citep{Papadopoulos2022}. As an illustration, adopting a fixed excitation temperature of $T_{\rm ex}=30$\,K would increase the inferred $M_{\rm H_2,[CI]}$ by a factor of about 1.5 relative to our fiducial assumption.

Dust–based masses, $M_{\rm H_2,dust}={\delta_{\rm GDR}}\times M_{\rm dust}$, are most sensitive to $\delta_{\rm GDR}$, the opacity law $\kappa_\nu$, and the dust temperature $T_{\rm d}$. 
At fixed observing frequency and flux density, $M_{\rm dust}\propto[\kappa_\nu B_\nu(T_{\rm d})]^{-1}$. Thus, increasing $\delta_{\rm GDR}$ linearly boosts $M_{\rm H_2}$, while higher $\kappa_\nu$ or higher $T_{\rm d}$ reduce $M_{\rm dust}$. 
In Figure \ref{fig:H2_mass_vs_LIR}, raising $T_{\rm d}$ from 47\,K to 100\,K lowers $M_{\rm H_2,dust}$ by $\approx0.6$\,dex. 
Adopting instead the \citet{daCunha2021} opacity prescription, $\kappa_\nu=\kappa_0(\nu/\nu_0)^{\beta}$ with $\kappa_0=0.77~{\rm cm^2\,g^{-1}}$ at $\nu_0=353$\,GHz and $\beta=1.9$, yields dust–based gas masses smaller by $\approx0.2$\,dex compared to our fiducial $\kappa_\nu$ assumption \citep{Decarli2022}.

For [C\,\textsc{ii}], beyond $T_{\rm ex,[CII]}$ and $X_{\rm C^+}$, the relevant uncertainty is the fraction of the 158\,$\mu$m emission that truly arises in molecular gas rather than from ionized or atomic phases. Methods that tie $\alpha_{[\mathrm{CII}]}$ to the [C\,\textsc{ii}] surface brightness and compactness \citep[e.g.,][]{Sommovigo2021} predict very low conversion factors ($\alpha_{[\mathrm{CII]}}\!\sim\!1$) for the most compact, intensely star--forming quasars, which would drive $M_{\rm H_2,[CII]}$ far below the other estimates. Such tension likely reflects the extrapolation of local scaling relations (Kennicutt--Schmidt and SFR--$L_{\rm [CII]}$) into regimes where the [C\,\textsc{ii}] emissivity is suppressed by grain charging (lower photoelectric heating efficiency), optical--depth/self--absorption, or partial thermalization in extreme radiation fields \citep[e.g.,][]{Herrera-Camus2018,Sutter2021}.

While the four tracers agree on average, at the highest $L_{\rm TIR}$ we find that the [C\,\textsc{ii}]–based masses tend to fall below the CO, [C\,\textsc{i}], and dust estimates. 
In luminosity space (see Figure \ref{fig:lum_linetoTIR} and  \ref{fig:lum_ciitoline}), the same objects exhibit (a) a declining $L_{\rm [CII]}/L_{\rm TIR}$ (the [C\,\textsc{ii}] deficit; e.g., \citealt{Lutz2016,Diaz-Santos2017,Herrera-Camus2018}), (b) a rising $L_{\rm [CII]}/L_{\rm [CI]}$ with $L_{\rm TIR}$, and (c) a roughly flat $L_{\rm [CII]}/L_{\rm CO(7-6)}$. 
A plausible interpretation is that, at high compactness and radiation–field intensity, grain charging reduces the photoelectric
heating efficiency in PDRs (\citealt{Bakes1994}; see also \citealt{Berne2022}) and large far–IR optical depths and/or [C\,\textsc{ii}] self–absorption further depress the [C\,\textsc{ii}] emissivity (\citealt{Goldsmith2012,Sutter2021}). Carbon is simultaneously driven from C$^{0}$ to C$^{+}$, raising $L_{\rm [CII]}/L_{\rm [CI]}$ (e.g., \citealt{Valentino2020}). Meanwhile, CO(7--6) emission primarily traces warm and dense molecular gas located deeper inside the clouds, and can be enhanced in compact star-forming regions where both gas temperature and density increase \citep{Lu2015,Kamenetzky2016a,Carilli2013}, keeping $L_{\rm [CII]}/L_{\rm CO(7-6)}$ roughly constant. If a single, fixed $\alpha_{[\mathrm{CII}]}$ is used, this reduced line emissivity per unit molecular mass naturally translates into an underestimate of $M_{\rm H_2,[CII]}$ at the bright end (see also \citealt{Narayanan2017,Sommovigo2021}), highlighting that the systematically lower [C II]-based gas masses reflects changing excitation/opacity, not a contradiction with the observed luminosity ratios.

Beyond the PDR-related effects discussed above, an important distinction among alternative heating mechanisms lies in their ability to heat dust relative to gas. While several non-radiative heating processes expected in compact starbursts---such as cosmic rays, turbulence, and shocks---can efficiently heat the molecular gas and modify carbon chemistry, they generally couple much more weakly to dust and therefore do not necessarily produce a commensurate increase in the FIR continuum \citep[e.g.,][]{Luhman2003,Papadopoulos2010,Meijerink2011}. As a result, these mechanisms alone do not naturally predict a systematic suppression of multiple gas cooling tracers when normalized by $L_{\rm TIR}$.

In contrast, AGN-related heating provides a more direct pathway to enhance the FIR continuum and dilute line-to-continuum ratios. Hard UV and X-ray radiation from the AGN can efficiently heat dust \citep{Sargsyan2012,Carilli2013}, while also altering the ionization and chemical balance of the ISM. In particular, high-energy photons can ionize C$^{+}$ to C$^{2+}$ and partially destroy CO, suppressing carbon-bearing lines and redistributing cooling to other fine-structure transitions such as [O\,\textsc{i}]~63~$\mu$m and [O\,\textsc{iii}]~88~$\mu$m \citep{Luhman2003,Abel2005,Langer2015}. This combination of enhanced dust heating and increased ionization offers a physically motivated explanation for the stronger decoupling between gas cooling tracers and $L_{\rm TIR}$ inferred for the most infrared-luminous sources, relative to starburst- and star-formation-dominated systems \citep{Greve2014,Herrera-Camus2018,Valentino2020a,Gururajan2023}. Future spectroscopic diagnostics---such as [N\,\textsc{ii}]\,122/205\,$\mu$m for the ionized fraction, [O\,\textsc{i}]/[O\,\textsc{iii}] for the cooling budget, [C\,\textsc{ii}] resolved profiles for optical depth, and mid–IR AGN tracers—will be crucial to test this scenario.

In summary, reasonable literature choices for $(\alpha_{\rm CO},r_{71},{\delta_{\rm GDR}},\kappa_\nu,T_{\rm d},T_{\rm ex,[CI]}, X_{\rm CI},T_{\rm ex,[CII}, X_{\rm C^+}$ can shift individual estimates by factors of a few; with our fiducial set, the four methods yield mutually consistent $M_{\rm H_2}$ within a factor of three. Breaking the remaining degeneracies will require spatially resolved [C\,\textsc{ii}] and [C\,\textsc{i}] to map phases and optical depths, and multi--$J$ CO to constrain excitation. Even so, the tracer cross-validation presented below provides a practical, internally consistent estimate of $M_{\rm H_2}$ for each source.

\begin{figure}[t]
  \centering
  \includegraphics[width=\linewidth]{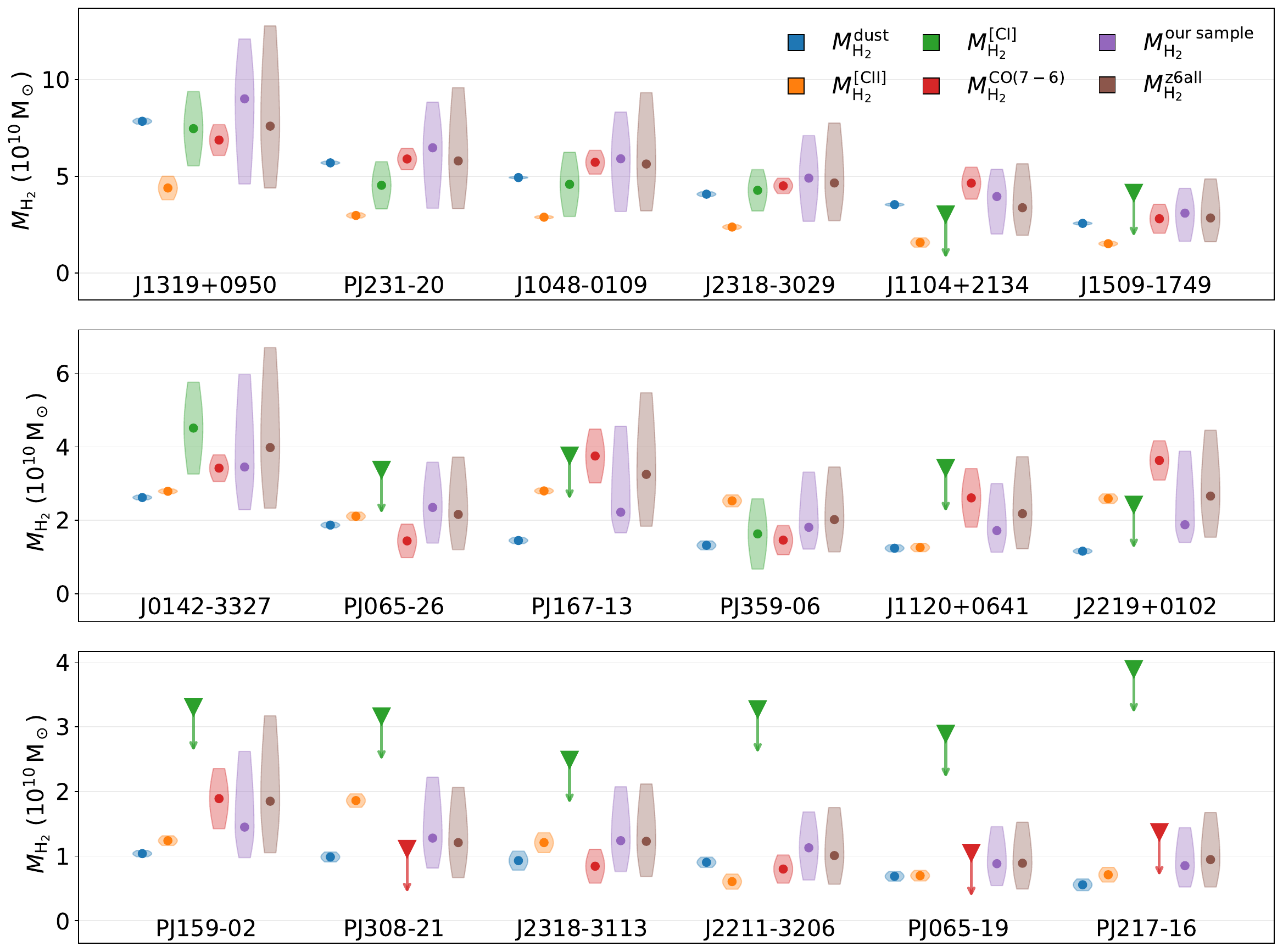}
  \caption{%
  Comparison of six molecular gas mass estimates for 18 sources, arranged in three horizontal panels (ordered from high to low $L_{\rm TIR}$, left to right and top to bottom). For each source, the estimates from $M_{\mathrm{H_2}}^{\mathrm{dust}}$, $M_{\mathrm{H_2}}^{\mathrm{[CII]}}$, $M_{\mathrm{H_2}}^{\mathrm{[CI]}}$, $M_{\mathrm{H_2}}^{\mathrm{CO(7\!-\!6)}}$, $M_{\mathrm{H_2}}^{\mathrm{our\; sample}}$, and $M_{\mathrm{H_2}}^{\mathrm{z6all}}$ are shown as colored points with spindle-shaped Gaussian error bars. Upper limits are indicated by downward arrows. For $M_{\mathrm{H_2}}^{\mathrm{our\; sample}}$ and $M_{\mathrm{H_2}}^{\mathrm{z6all}}$, asymmetric uncertainties are used; the other four masses use symmetric errors. Masses and errors are scaled by $10^{10}\,M_\odot$. 
  }
  \label{fig:mh2_comparison}
\end{figure}

\subsubsection{Hierarchical cross–calibration of molecular–gas tracers}
\label{sec:xcal}

Our goal is to infer molecular gas masses by jointly exploiting four tracers---CO(7--6), [C\,\textsc{i}](2--1), [C\,\textsc{ii}], and the (sub)millimeter dust continuum—and to cross–calibrate their conversion factors in a single, self–consistent framework. As outlined in Figure \ref{fig:bayes_flow}, we adopt a hierarchical Bayesian model in which each observable is written as a linear relation in log space between a latent gas mass and a tracer–specific coefficient, with intrinsic scatter and full propagation of measurement uncertainties (including upper limits via censored likelihoods):
\begin{align}
\log L_{\rm CO(7-6)}      &= \log M_{\rm H_2} + \log k_{\rm CO},\\
\log L_{\rm [CI](2-1)}    &= \log M_{\rm H_2} + \log k_{\rm [CI]},\\
\log L_{\rm [CII]}        &= \log M_{\rm H_2} + \log k_{\rm [CII]},\\
\log S_{\nu,\,{\rm mJy}}  &= \log M_{\rm H_2} + \log k_{\rm dust}.
\end{align}
For CO(7--6) we reparameterize $\log k_{\rm CO}=A_{\rm CO}+\log r_{71}-\log\alpha_{\rm CO}$. $A_{\rm CO}\!=\!\log\!\big[3\times10^{-11}\,\nu_{\rm CO(7-6)}^3\big]$ (with $\nu$ in gigahertz) converts brightness--temperature luminosity $L'$ to $L$. 
For the fine–structure lines we tie the level populations to excitation temperatures through the partition functions, $\log k_{\rm [CI]}=\log Q_{21}(T_{\rm ex,[CI]})+\log X_{\rm CI}-C_{\rm [CI]}$ and $\log k_{\rm [CII]}=\log Q_{\rm [CII]}(T_{\rm ex,[CII]})+\log X_{\rm C^+}-C_{\rm [CII]}$. $C_{\rm [CI]}$ and $C_{\rm [CII]}$ are tracer--dependent constants that gather the Einstein coefficients, level degeneracies, and the $L$--$L'$ conversion (including the $\nu^3$ term). 
For the dust continuum, we decompose the coefficient into a source-level term accounting for geometry and the CMB and a population-level term describing the ISM physics,
\begin{align}
k_{\rm dust} &\;= \log k_{\rm dust, geom}+ \log k_{\rm dust, phys} - \log{\delta_{\rm GDR}},\\
k_{\rm dust,geom} &\;\equiv\; \frac{(1+z)\,\big[B_\nu(T_{\rm d})-B_\nu\!\big(T_{\rm CMB}(z)\big)\big]}{D_L^2},\\
k_{\rm dust,phys} &\;\equiv\; 10^{26}\,\kappa_\nu\,\frac{1-e^{-\tau_\nu}}{\tau_\nu}\,M_\odot[{\rm g}],
\end{align}
so that $S_\nu({\rm mJy}) = M_{\rm H_2}\,k_{\rm dust,geom}\,k_{\rm dust,phys}/{\delta_{\rm GDR}}$. The factor $10^{26}$ converts cgs flux density to mJy, and $M_\odot[{\rm g}]$ converts the model’s gas mass (in $M_\odot$)
to grams inside $k_{\rm dust,phys}$. The rest frequency $\nu$ is taken at the continuum reference used in the fit
(e.g., \ $\nu\!\approx\!1900.5$\,GHz for 158\,$\mu$m). We also allowed a weak population–level trend (i.e., a global trend across the sample) of gas mass with rest–UV luminosity via
\begin{equation}
\log M_{\rm H_2} \sim \mathcal{N}\big(\mu_M + \beta_M\,[M_{1450}-\tilde{M}_{1450}],\,\sigma_{\rm src}^2\big),
\end{equation}
where $\tilde M_{1450}$ is a fixed pivot (here set to -25.0) that reduces covariance between slope and intercept, $\mu_M$ is the population mean at the pivot, $\beta_M$ (in dex per mag) is the slope linking $M_{1450}$ to $\log M_{\rm H_2}$, and $\sigma_{\rm src}$ is the intrinsic, source–to–source scatter beyond measurement uncertainties. This term simultaneously acts as a hierarchical prior that partially pools individual sources toward the population trend and as a regression that infers the strength of the $M_{\rm H_2}$–$M_{1450}$ correlation via $\beta_M$. The formulation is readily extensible to additional covariates (e.g., dynamical mass or size) in the future.

Posterior sampling was performed with the No–U–Turn Sampler (NUTS; \citealt{Hoffman2011}). Upper limits are incorporated through the normal cumulative distribution function (CDF) term in the log–likelihood. We report medians and 68\% highest–density intervals (HDIs) for global parameters (Table~\ref{tab:xcal_post}) and provide per–source masses in Table~\ref{tab:mass}.

Figure~\ref{fig:mh2_comparison} stacks, for each of our 18 quasars (ordered from high to low $L_{\rm TIR}$, left to right and top to bottom), the molecular-gas masses inferred from the four ``classical'' tracers (dust, [C\,\textsc{ii}], [C\,\textsc{i}], CO) alongside the hierarchical posteriors obtained when conditioning on (i) only the 18–object fit ($M_{\rm H_2,18}$) and (ii) the \texttt{Allz6quasar} sample fit applied to these same 18 objects ($M_{\rm H_2,z6}$). Overall, single–tracer masses agree within a factor $\lesssim3$, with [C\,\textsc{i}] showing the largest spread, primarily due to its lower observational S/N, and [C\,\textsc{ii}] tending slightly low for the most IR–luminous sources. In contrast, the median value of the hierarchical masses lies between the single–tracer estimates and exhibits noticeably larger uncertainties. This is because the hierarchical approach accounts for the scatter among the single–tracer estimates. It also adopts Gaussian priors on all global parameters, rather than fixing them directly as in the single–tracer case, thereby leading to broader uncertainties. For all 18 sources, $M_{\rm H_2,18}$ and $M_{\rm H_2,z6}$ remain essentially indistinguishable.

Table~\ref{tab:xcal_post} compares the posteriors from our hierarchical fit for (i) our 18–object subsample and (ii) the full $z\sim6$ quasar sample. The four calibration coefficients are mutually consistent between the two runs: $\log k_{\rm CO}$, $\log k_{\rm [CI]}$, $\log k_{\rm [CII]}$, and the dust factor $\log k_{\rm dust,phys}$ agree within the 68\% HDIs. The underlying conversion/excitation parameters are likewise in agreement: $\alpha_{\rm CO}\approx0.8$, $r_{71}\approx0.17$, $T_{\rm ex,[CI]}\sim47$\,K, $T_{\rm ex,[CII]}\sim100$\,K, $\log X_{\rm CI}\simeq-4.8$, $\log X_{\rm C^+}\simeq-4.4$, implying $\alpha_{\rm [CII]}\sim14~M_\odot\,L_\odot^{-1}$. Gas--to--dust ratios are consistent ($\delta_{\rm GDR}\sim100$), while intrinsic scatters show small but interpretable differences: our 18–object run yields a lower dust scatter (0.12\,dex versus 0.22\,dex), plausibly reflecting more homogeneous continuum data, whereas the full sample shows a slightly larger source–to–source dispersion ($\sigma_{\rm src}=0.39$\,dex versus 0.30\,dex). The mass–AGN regression is weak in the 18–object fit ($\beta_M=-0.03^{+0.08}_{-0.06}\,\mathrm{dex\,mag^{-1}}$) but becomes mildly negative in the full compilation ($\beta_M\approx-0.12\,\mathrm{dex\,mag^{-1}}$), suggesting that more UV–luminous quasars tend to host somewhat larger $M_{\rm H_2}$ over the broader $M_{1450}$ range.

In Table \ref{tab:xcal_comb}, we compare the six cross–calibration combinations defined by \citet[][their Eqs. 8–13]{Gururajan2023} with the posteriors from our hierarchical fit. Relative to their lensed DSFG sample, our medians for $X_{\rm CI}\alpha_{\rm CO}$, $X_{\rm CI}\alpha_{\rm [CII]}$, $X_{\rm CI}\,\delta_{\rm GDR}$, $\alpha_{\rm CO}/\alpha_{\rm [CII]}$, $\delta_{\rm GDR}/\alpha_{\rm CO}$, and $\delta_{\rm GDR}/\alpha_{\rm [CII]}$ show lower $X_{\rm CI}\alpha_{\rm CO}$ and $X_{\rm CI}\alpha_{\rm [CII]}$ by factors of a few whereas the ratio $\alpha_{\rm CO}/\alpha_{\rm [CII]}$ is consistent within uncertainties, and the combinations with $\delta_{\rm GDR}$ in the numerator ($\delta_{\rm GDR}/\alpha_{\rm CO}$, $\delta_{\rm GDR}/\alpha_{\rm [CII]}$) are correspondingly higher. Some ingredients plausibly drive the offsets. (i) Population and redshift: \citet{Gururajan2023} focus on lensed DSFGs at $z\!\sim\!2\!-\!5$, whereas we fit luminous quasar hosts at $z\!\sim\!6$, where the higher CMB background and compact, high–$\Sigma_{\rm IR}$ star formation lower the observed dust/line contrast and tend to favor slightly higher $\delta_{\rm GDR}$ and lower $\alpha_{\rm [CII]}$, consistent with the [C\,\textsc{ii}] deficit. (ii) [C\,\textsc{i}] treatment: they calibrate with [C\,\textsc{i}](1–0) through $Q_{10}$, while we place [C\,\textsc{i}](2–1) directly in the likelihood via $Q_{21}(T_{\rm ex,[CI]})$ and fit $T_{\rm ex,[CI]}$ jointly; this choice tends to pull $X_{\rm CI}\alpha_{\rm CO}$ and $X_{\rm CI}\alpha_{\rm [CII]}$ slightly lower at fixed [C\,\textsc{i}] luminosity. Crucially, running the same hierarchical model on (a) the full $z \sim 6$ quasar compilation and (b) our 18–object subsample yields nearly identical medians for all six combinations, indicating that these inferences are robust to sample selection and are not driven by a handful of sources.

\begin{table*}
\centering
\caption{Derived molecular gas masses. \label{tab:mass}}
\begin{tabular}{lcccccc}
\hline
Target 
& $M_{\rm H_2}^{\rm dust}$ 
& $M_{\rm H_2}^{\rm [CII]}$ 
& $M_{\rm H_2}^{\rm [CI]}$ 
& $M_{\rm H_2}^{\rm CO(7-6)}$ 
& $M_{\rm H_2}^{\rm our\; sample}$ 
& $M_{\rm H_2}^{\rm z6all}$ \\
& [$\rm 10^{10}\, M_\odot$] & [$\rm 10^{10}\, M_\odot$] & [$\rm 10^{10}\, M_\odot$] 
& [$\rm 10^{10}\, M_\odot$] & [$\rm 10^{10}\, M_\odot$] & [$\rm 10^{10}\, M_\odot$] \\
\hline
J1120+0641 & $1.24 \pm 0.10 $ & $1.26 \pm 0.11 $ & $<3.42 $ & $2.61 \pm 0.80 $ & $1.72 \pm 0.93 $ & $2.18 \pm 1.25 $ \\
J1104+2134 & $3.54 \pm 0.07 $ & $1.58 \pm 0.25 $ & $<3.03 $ & $4.65 \pm 0.83 $ & $3.96 \pm 1.68 $ & $3.38 \pm 1.85 $ \\
J1048-0109 & $4.94 \pm 0.05 $ & $2.89 \pm 0.08 $ & $4.59 \pm 1.66 $ & $5.73 \pm 0.61 $ & $5.91 \pm 2.57 $ & $5.64 \pm 3.05 $ \\
PJ231-20 & $5.70 \pm 0.08 $ & $2.98 \pm 0.14 $ & $4.54 \pm 1.22 $ & $5.90 \pm 0.55 $ & $6.48 \pm 2.75 $ & $5.80 \pm 3.13 $ \\
PJ167-13 & $1.45 \pm 0.08 $ & $2.80 \pm 0.08 $ & $<3.76 $ & $3.75 \pm 0.73 $ & $2.22 \pm 1.45 $ & $3.25 \pm 1.81 $ \\
J2318-3113 & $0.93 \pm 0.15 $ & $1.21 \pm 0.15 $ & $<2.49 $ & $0.84 \pm 0.26 $ & $1.24 \pm 0.66 $ & $1.23 \pm 0.72 $ \\
PJ159-02 & $1.04 \pm 0.05 $ & $1.24 \pm 0.08 $ & $<3.30 $ & $1.89 \pm 0.47 $ & $1.45 \pm 0.82 $ & $1.85 \pm 1.06 $ \\
J2211-3206 & $0.91 \pm 0.08 $ & $0.61 \pm 0.12 $ & $<3.27 $ & $0.80 \pm 0.22 $ & $1.12 \pm 0.53 $ & $1.01 \pm 0.59 $ \\
J0142-3327 & $2.62 \pm 0.06 $ & $2.79 \pm 0.06 $ & $4.51 \pm 1.25 $ & $3.42 \pm 0.36 $ & $3.45 \pm 1.84 $ & $3.98 \pm 2.19 $ \\
PJ308-21 & $0.99 \pm 0.08 $ & $1.86 \pm 0.10 $ & $<3.15 $ & $<1.11 $ & $1.27 \pm 0.70 $ & $1.21 \pm 0.70 $ \\
PJ065-26 & $1.87 \pm 0.08 $ & $2.11 \pm 0.11 $ & $<3.37 $ & $1.44 \pm 0.46 $ & $2.35 \pm 1.10 $ & $2.16 \pm 1.26 $ \\
PJ359-06 & $1.32 \pm 0.12 $ & $2.53 \pm 0.16 $ & $1.63 \pm 0.95 $ & $1.46 \pm 0.40 $ & $1.81 \pm 1.05 $ & $2.02 \pm 1.16 $ \\
PJ217-16 & $0.56 \pm 0.09 $ & $0.71 \pm 0.11 $ & $<3.89 $ & $<1.37 $ & $0.85 \pm 0.46 $ & $0.95 \pm 0.57 $ \\
J2219+0102 & $1.16 \pm 0.07 $ & $2.59 \pm 0.13 $ & $<2.42 $ & $3.63 \pm 0.53 $ & $1.88 \pm 1.24 $ & $2.66 \pm 1.46 $ \\
J2318-3029 & $4.08 \pm 0.12 $ & $2.38 \pm 0.12 $ & $4.28 \pm 1.07 $ & $4.51 \pm 0.39 $ & $4.91 \pm 2.21 $ & $4.66 \pm 2.53 $ \\
J1319+0950 & $7.85 \pm 0.15 $ & $4.40 \pm 0.61 $ & $7.47 \pm 1.92 $ & $6.88 \pm 0.80 $ & $9.01 \pm 3.75 $ & $7.60 \pm 4.19 $ \\
PJ065-19 & $0.69 \pm 0.07 $ & $0.70 \pm 0.08 $ & $<2.89 $ & $<1.05 $ & $0.88 \pm 0.46 $ & $0.89 \pm 0.52 $ \\
J1509-1749 & $2.57 \pm 0.07 $ & $1.52 \pm 0.12 $ & $<4.14 $ & $2.81 \pm 0.75 $ & $3.10 \pm 1.37 $ & $2.85 \pm 1.63 $ \\\hline
\end{tabular}
\end{table*}

\begin{table*}
\centering
\caption{Cross--calibration parameter combinations derived from our posterior samples (medians with 68\% HDIs) compared to \citet{Gururajan2023}}
\label{tab:xcal_comb}

\begin{tabular}{lccc}
\hline
Combination & Our sample & All quasars & Gururajan+23 \\
\hline
$X_{\rm CI}\,\alpha_{\rm CO}$ ($\times10^{-5}$) & $1.26\,[0.53,\,2.99]$ & $1.45\,[0.61,\,3.47]$ & $6.31\pm0.67$ \\
$X_{\rm CI}\,\alpha_{\rm [CII]}$ ($\times10^{-5}$) & $21.0\,[10.7,\,41.6]$ & $24.8\,[11.9,\,51.8]$ & $95.5\pm17.1$ \\
$X_{\rm CI}\,\delta_{\rm GDR}$ ($\times10^{-5}$) & $159\,[81.6,\,311]$ & $216\,[102,\,465]$ & $302\pm52$ \\
$\alpha_{\rm CO}/\alpha_{\rm [CII]}$ & $0.060\,[0.026,\,0.139]$ & $0.059\,[0.025,\,0.139]$ & $0.08\pm0.01$ \\
$\delta_{\rm GDR}/\alpha_{\rm CO}$ & $126\,[54.6,\,290]$ & $149\,[61.5,\,366]$ & $42.7\pm6.4$ \\
$\delta_{\rm GDR}/\alpha_{\rm [CII]}$ & $7.55\,[3.99,\,14.3]$ & $8.72\,[4.15,\,18.5]$ & $4.36\pm1.07$ \\
\hline
\end{tabular}
\\[2pt]
\tablefoot{Our values are derived from posterior samples of the hierarchical fit (columns 2--3); numbers in brackets are 68\% HDIs. \citet{Gururajan2023} report fiducial values with 1$\sigma$ uncertainties.}
\end{table*}

\begin{figure*}
    \centering
    \includegraphics[width=0.3\textwidth]{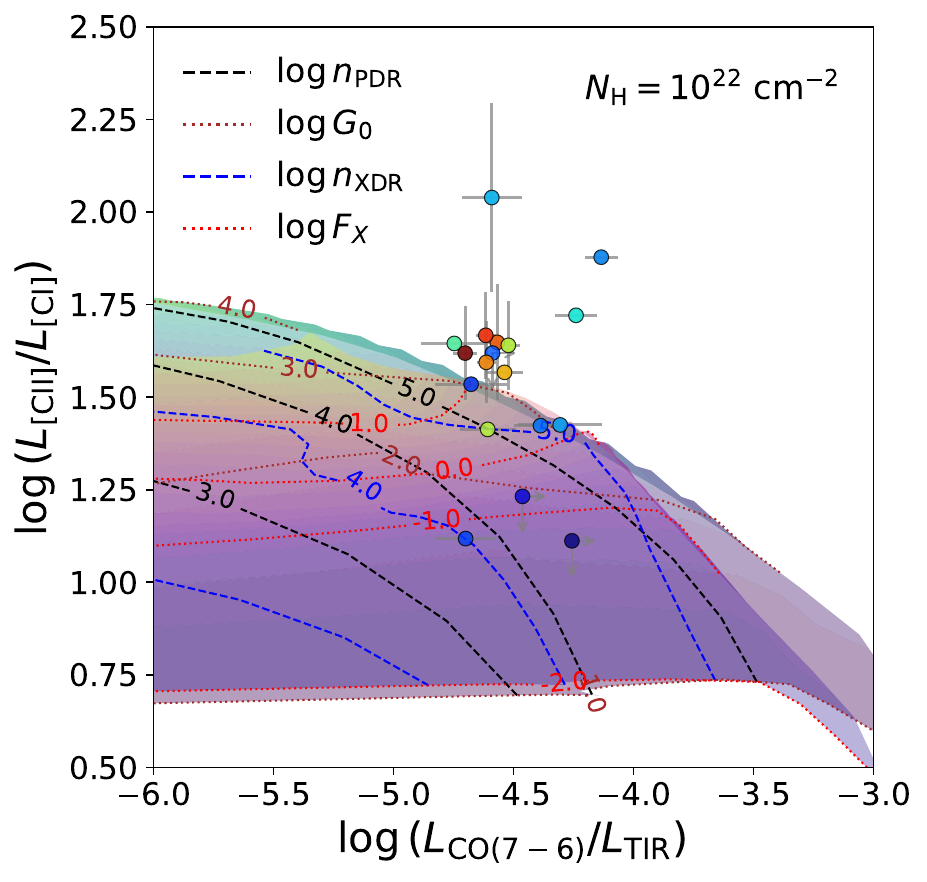}
    \includegraphics[width=0.283\textwidth]{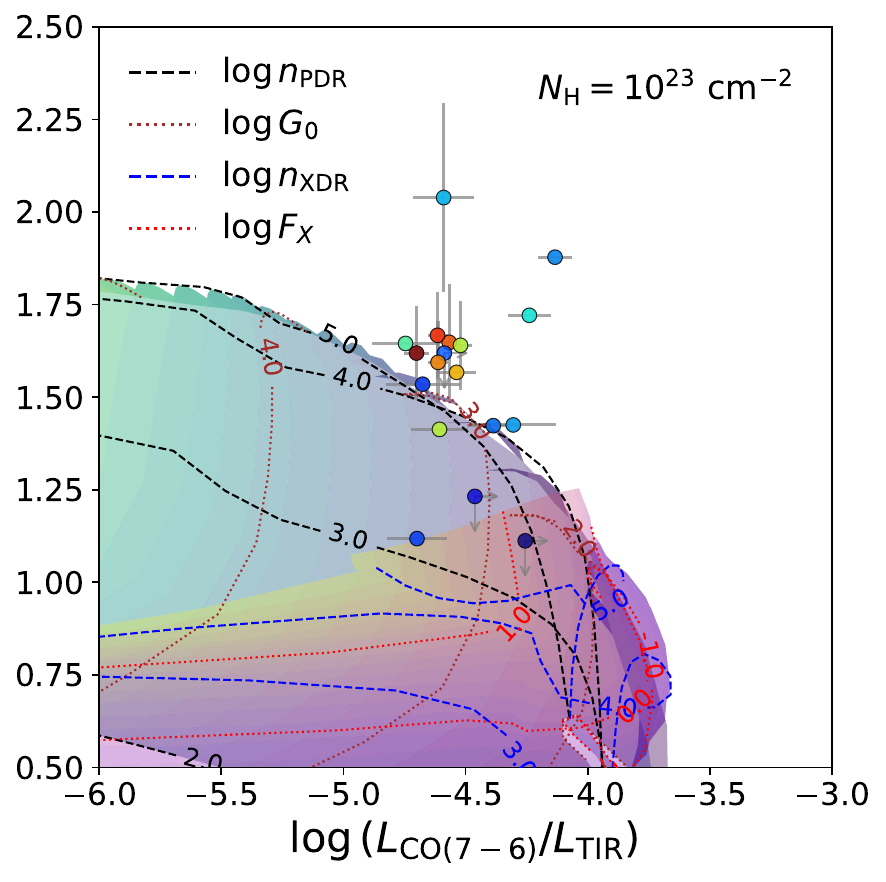}
    \includegraphics[width=0.38\textwidth]{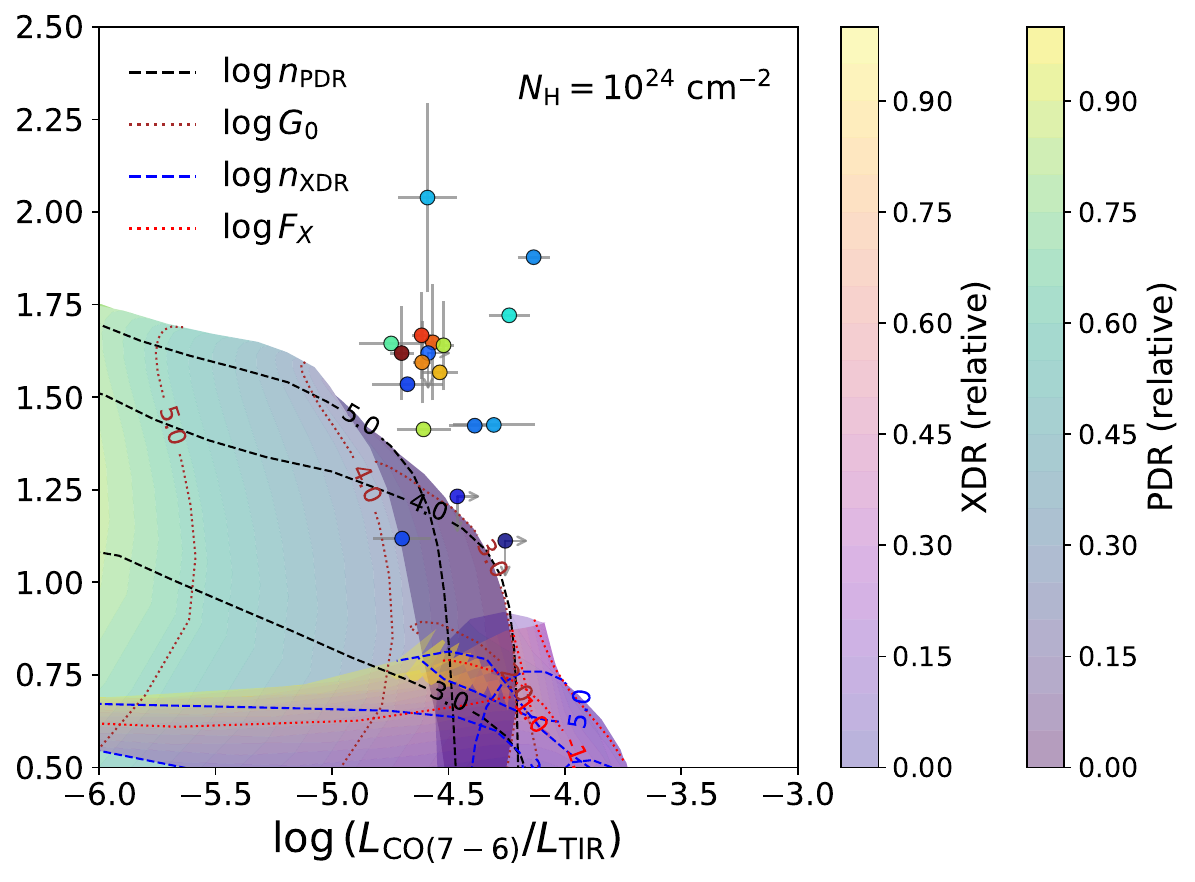}
    \caption{
        Diagnostic diagrams comparing observed luminosity ratios of [C\,\textsc{ii}], [C\,\textsc{i}], and CO(7–6) with PDR and XDR model predictions under different column densities. 
        From left to right, the panels correspond to $N_{\mathrm{H}} = 10^{22}$, $10^{23}$, and $10^{24}~\mathrm{cm}^{-2}$, respectively. Background color maps show model-predicted $\log L_{\mathrm{TIR}}$ values normalized to the maximum within each grid, illustrating relative variations across the parameter space, with PDR and XDR parameter grids overlaid. Contours represent constant gas density ($\log n$) and radiation strength ($\log G_0$ for PDR and $\log F_X$ for XDR). We note that PDR models probe FUV-heated cloud surfaces and may not represent the average ISM condition.
        Observational points are color coded by $M_{1450}$ or $L_{\mathrm{TIR}}$, with associated uncertainties and upper-limit arrows where applicable.
        The color of observed data points is consistent with previous figures and represents $L_{\mathrm{TIR}}$. 
    }
    \label{fig:pdr_xdr_NH_series}
\end{figure*}

\begin{figure}
    \centering
    \includegraphics[width=\hsize]{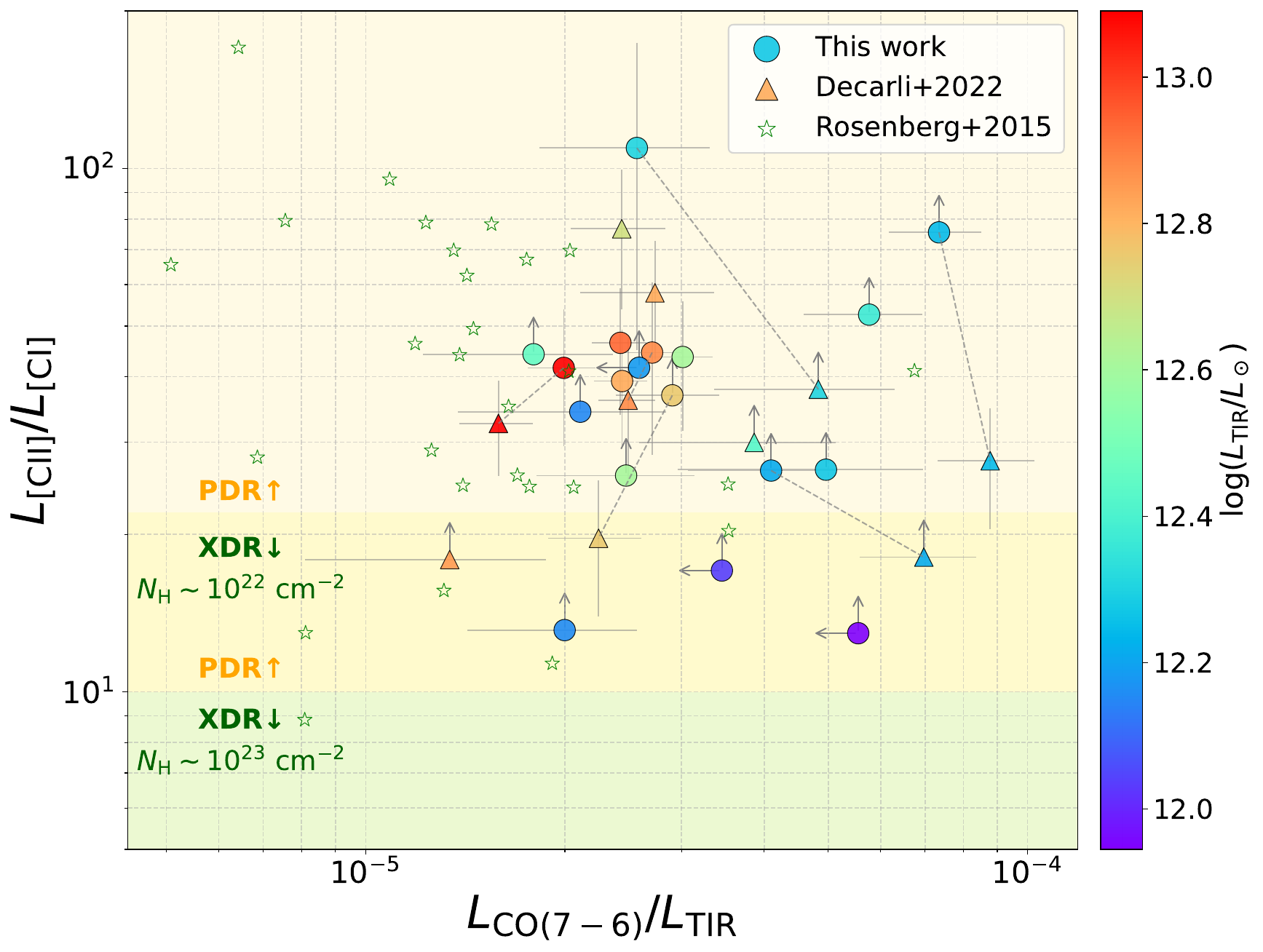}
    \caption{
    Diagnostic diagram comparing $L_{\mathrm{[CII]}}/L_{\mathrm{[CI]}}$ and $L_{\mathrm{CO(7-6)}}/L_{\mathrm{TIR}}$ for high-redshift quasar host galaxies.
    The filled circles represent the sources presented in this work.
    Triangles indicate the $z > 6$ quasar sample from \citet{Decarli2022} based on NOEMA observations. 
    Two sources connected by dashed lines denote multiple measurements of the same object. Green hollow pentagrams show local galaxies from \citet{Rosenberg2015}.
    The shaded regions indicate the predicted transitions between PDR- and XDR-dominated regimes, based on the column densities of $N_{\mathrm{H}} = 10^{22}$ (golden shaded) and $10^{23}$ cm$^{-2}$ (light-green shaded), following the models of \citet{Decarli2022}.
    }
    \label{fig:cii_ci_co_diag}
\end{figure}

\subsection{Physical properties of the cold ISM} \label{sec:diagnostics}

To investigate the excitation conditions of the ISM in our sample of $z \gtrsim 6$ quasar host galaxies, we compare the observed line luminosity ratios—$L_{\rm [CII]}/L_{\rm [CI]}$ and $L_{\rm CO(7\text{--}6)}/L_{\rm TIR}$—with predictions from PDR and XDR models. Figure~\ref{fig:pdr_xdr_NH_series} displays three diagnostic diagrams corresponding to different assumed hydrogen column densities, $N_{\mathrm{H}} = 10^{22}$, $10^{23}$, and $10^{24}\,\mathrm{cm^{-2}}$. The models are based on the grids developed by \citet{Pensabene2021}, and include predictions for the gas density ($\log n/{\rm cm^{-3}}$) and either the FUV radiation field strength ($\log G_0$, in Habing units) for PDRs or the incident X-ray flux ($\log F_X/{\rm erg\,s^{-1}\,cm^{-2}}$) for XDRs.

The colored backgrounds represent the normalized total infrared luminosity ($L_{\mathrm TIR}$) predicted by each model grid, providing an additional, albeit model-dependent, diagnostic dimension. Overlaid contours correspond to constant gas densities and radiation field strengths. Observed sources are plotted with their line ratios (see Table~\ref{tab:mass}), with marker color indicating $L_{\mathrm TIR}$ (as in previous figures).

We find that only a subset of the observed ratios falls within the parameter space covered by the single–zone PDR and XDR model grids. A small number of sources are consistent with PDR-like conditions, clustering around $\log n/{\rm cm^{-3}} \!\sim\! 3$–$5$ and $\log G_0 \!\sim\! 3$–$4$ at moderate column densities ($N_{\mathrm H} \sim 10^{23}\,\mathrm{cm^{-2}}$). We note, however, that sustaining such high $G_0$ values as an average property of the molecular ISM may be challenging in dusty environments due to attenuation of FUV photons by dust within and between molecular clouds \citep{Wolfire1990}. Moreover, the majority of objects lie outside the regions predicted by either model, particularly at the high column density ($N_{\mathrm H} = 10^{24}\,\mathrm{cm^{-2}}$), where the observed $L_{\rm [CII]}/L_{\rm [CI]}$ ratios exceed the range expected from neutral or molecular gas excitation within a single PDR or XDR component. This systematic offset indicates that single–zone PDR or XDR models do not provide a complete description of the integrated ISM conditions in these systems.

Overall, the results point to a diversity of ISM excitation conditions among the high-redshift quasars. Although a subset of sources shows ratios broadly consistent with PDR-like excitation, in line with previous studies \citep[e.g.,][]{Decarli2022}, the majority cannot be reproduced by a single PDR component. As discussed in Section~\ref{sec:gas_mass}, additional non-radiative heating processes—most notably mechanical heating associated with turbulence and shocks, as well as enhanced cosmic-ray heating—may also contribute to the observed line ratios. Such mechanisms are expected to heat the molecular gas efficiently without proportionally increasing the dust temperature, and therefore are unlikely to fully account, on their own, for the pronounced deficits observed in $L_{\rm [CII]}/L_{\rm TIR}$, $L_{\rm [CI]}/L_{\rm TIR}$, and $L_{\rm CO(7\text{--}6)}/L_{\rm TIR}$. However, they can act in combination with radiative heating in PDRs or XDRs and modify the excitation and chemical balance of the gas.

In particular, models including mechanical heating demonstrate that, at high gas densities and intense radiation fields (e.g., $\log n\,(\mathrm{cm^{-3}}) \sim 5.5$
 and high $\log G_0 \sim 5$), increasing the fractional contribution of mechanical energy dissipation can lead to elevated $L_{\rm [CII]}/L_{\rm [CI]}$ ratios relative to pure PDR predictions \citep{Kazandjian2015}. Similarly, enhanced cosmic-ray ionization rates can alter the carbon chemistry and gas excitation, affecting the relative strengths of [C\,\textsc{ii}], [C\,\textsc{i}], and CO emission \citep[e.g.,][]{Papadopoulos2010,Meijerink2011}. In the absence of spatially resolved data or a more complete set of diagnostic lines, these processes cannot be uniquely disentangled from UV- or X-ray–driven heating, underscoring that the interpretation of the observed line ratios likely requires a composite excitation scenario rather than a single dominant mechanism.

In Figure~\ref{fig:cii_ci_co_diag} we also compare, for each source in our sample, the observed ratios $L_{\rm [CII]}/L_{\rm [CI]}$ and $L_{\rm CO(7\text{--}6)}/L_{\rm IR}$ with those of the $z\!\sim\!6$ quasar sample from \citet{Decarli2022}, nearby galaxies compiled by \citet{Rosenberg2015}, and with the average model predictions shown in Figure~\ref{fig:pdr_xdr_NH_series}. For sources in the \citet{Decarli2022} sample that adopted literature $L_{\rm TIR}$ values, we recompute $L_{\rm TIR}$ following Section~\ref{sec:luminosity} to ensure a uniform comparison. Our quasars occupy a relatively narrow locus: all have $L_{\rm [CII]}/L_{\rm [CI]} > 20$, consistent with most high–$z$ sources and with the bulk of the local sample in \citet{Rosenberg2015}. For $L_{\rm CO(7\text{--}6)}/L_{\rm IR}$ we find values of $10^{-5}$–$10^{-4}$, in line with the other high–$z$ quasars and, on average, a factor of $\sim$2–3 higher than typical local galaxies in \citet{Rosenberg2015}. Given the different redshift regimes (and associated selection effects), we cannot determine whether this offset is primarily due to AGN contributions or to broader evolutionary differences.

\begin{figure}[t]
\centering
\includegraphics[width=\linewidth]{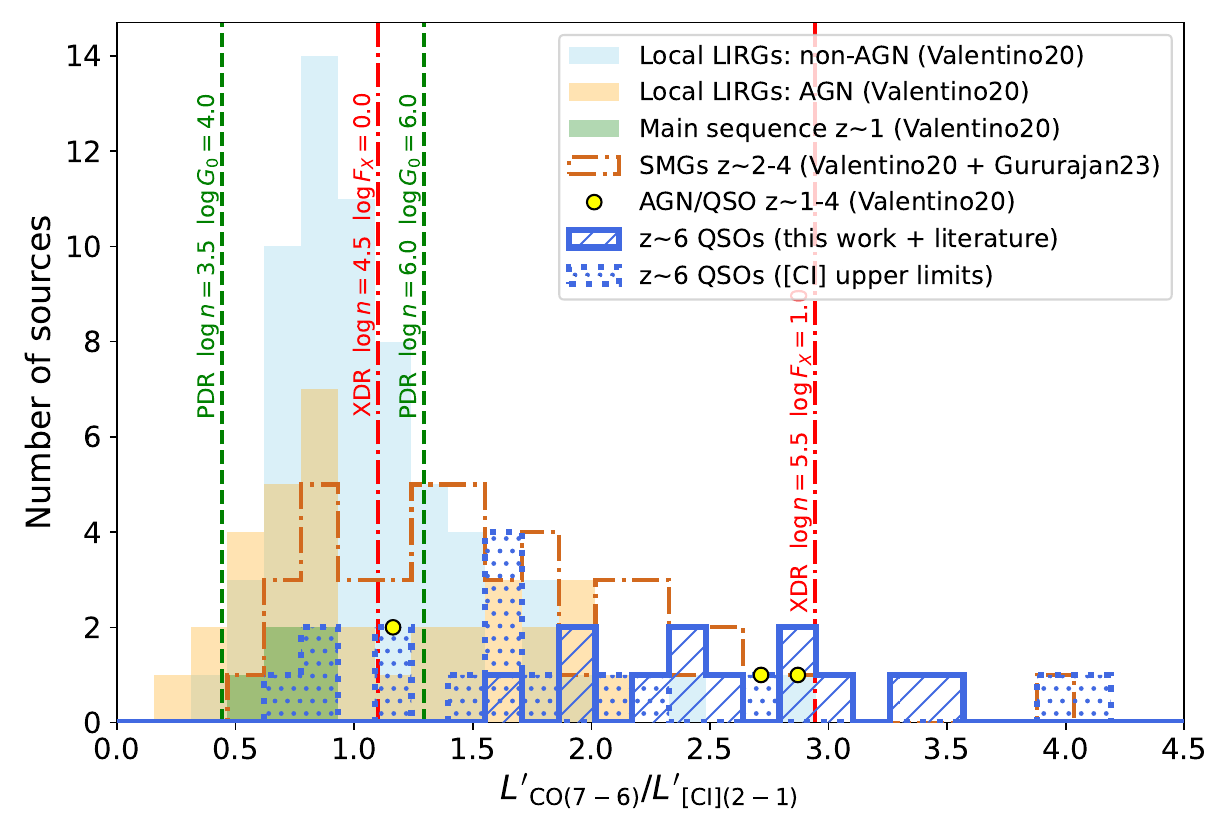}
\caption{
Distribution of the line luminosity ratio $L'_{\rm CO(7-6)}/L'_{\rm [CI](2-1)}$ 
for different galaxy populations. Local LIRGs from \citet{Valentino2020} are shown as filled histograms (sky-blue for non--AGN and orange for AGN). $z\sim1$ main-sequence galaxies are shown as green filled histograms, and SMGs at $z\sim2$--4 (from \citealt{Valentino2020} and \citealt{Gururajan2023}) as brown dash-dotted histograms. AGN/QSOs at $z\sim1$--4 are indicated by yellow circles. The $z\sim6$ quasars (this work and literature) are shown as blue hatched histograms, while sources with [C\,\textsc{i}](2--1) upper limits are shown as dotted and hatched blue histograms. Vertical green dashed lines indicate representative PDR model ratios, and dash-dotted red lines indicate representative XDR model ratios, both computed for a characteristic column density of $N_{\rm H}=10^{23}\,\mathrm{cm^{-2}}$. We note PDR predictions primarily probe cloud surfaces and may not represent the average conditions of the molecular ISM.
}
\label{fig:co76_ci21_ratio}
\end{figure}

The $L'_{\rm CO(7-6)}/L'_{\rm [CI](2-1)}$ ratio can be interpreted as an empirical proxy for the fraction of warm molecular gas relative to the total molecular gas reservoir, as emphasized by \citet{Andreani2018}. Figure~\ref{fig:co76_ci21_ratio} compares the distribution of this ratio for our $z\sim6$ quasar hosts with several reference samples from the literature, including local luminous infrared galaxies (LIRGs), $z\sim1$ main-sequence galaxies and $z\sim1-4$ quasi-stellar objects (QSOs)/AGN from the compilation \citet{Valentino2020}, as well as SMGs at $z\sim2$--4 compiled from \citet{Valentino2020} and \citet{Gururajan2023}. The vertical lines indicate representative ratios predicted by the PDR and XDR models discussed in the previous discussion. We note that the PDR predictions may trace only part of the molecular ISM conditions, as they primarily probe the surface layers of molecular clouds. In particular, main-sequence galaxies cluster around $L'_{\rm CO(7-6)}/L'_{\rm [CI](2-1)} \lesssim 1$, consistent with molecular gas heating dominated by classical far-UV PDRs. The local LIRG population, both AGN and non--AGN systems, largely occupies a similar regime, although the AGN LIRGs exhibit a slightly extended high-ratio tail and a weak bimodal structure. The SMG population and some luminous LIRGs extend to higher ratios that cannot be easily reproduced by standard PDR conditions alone. In these systems, additional volumetric heating sources associated with intense starbursts—such as enhanced cosmic-ray fields or mechanical heating from shocks and turbulence—may contribute to the elevated excitation. We note that many of the SMG population may also be affected by gravitational lensing, which can introduce additional uncertainties in the measured ratios. The Cloverleaf quasar at $z = 2.558$ shows an exceptionally high ratio of 9.16, likely reflecting extreme gas excitation and possible differential lensing effects \citep{Bradford2009}. As it strongly deviates from the rest of the sample, it is omitted from the plot for clarity.

In contrast, the $z\sim6$ quasars systematically occupy the high-ratio end of the distribution. Their detection ratios are significantly larger than those observed in local non--AGN LIRGs and main-sequence galaxies, and extend beyond the typical range of SMGs. A two-sample K–S test confirms that the $z\sim6$ quasar distribution differs strongly from local non--AGN LIRGs ($D=0.90$, $p\approx1.7\times10^{-9}$) and from main-sequence galaxies ($D=1.00$, $p\approx 3.2\times10^{-4}$). The  difference with SMGs is smaller but still significant ($D=0.63$, $p\approx 4.8\times10^{-4}$). These elevated ratios imply that a large fraction of the molecular gas in these early quasar hosts is in a highly excited phase. The systematically high values observed here are therefore consistent with a scenario in which additional heating sources, most naturally X-ray irradiation associated with the rapidly accreting AGN, play an important role in regulating the thermal state of the molecular ISM in these systems.

\section{Conclusions} \label{sec:conclusions}

We have presented a systematic ALMA survey targeting the CO(7--6) and [C\,\textsc{i}](2--1) emission lines in a sample of 18 quasars at $z \sim 6$, as part of project 2019.1.00147.S. This work significantly expands on previous efforts by increasing both the sensitivity and sample size relative to earlier NOEMA observations. Our main findings are summarized below:

\begin{itemize}
    \item CO(7--6) emission is detected in 15 out of 18 quasars, [C\,\textsc{i}](2--1) in 6 out of 18 quasars, and the $3$\,mm dust continuum in 13 out of 18 quasars. With these observations, the number of $z \sim 6$ quasars with CO(7--6) and/or [C    \,\textsc{i}] detections has significantly increased compared to previous studies, expanding the available sample for probing the molecular and atomic gas properties in the early Universe.
    
    \item We measure CO(7--6) and [C\,\textsc{i}](2--1) luminosities and derive $L_{\rm TIR}$ from the dust continuum, while the [C\,\textsc{ii}] luminosities are compiled from the literature. Across our sample, line luminosities increase with $L_{\rm TIR}$ but show no clear trend with $M_{1450}$. In line-to-continuum space we recover the classical [C\,\textsc{ii}] deficit, with many sources at or below the high–$\Sigma_{\rm SFR}$ locus.  CO(7--6)/TIR and [C\,\textsc{i}]/TIR also show analogous decreasing trends, which is not observed in main-sequence galaxies and starbursts. This is consistent with compact, intense star formation and/or enhanced dust opacity (with a possible AGN continuum contribution).

    \item We derive the molecular gas masses from dust, CO(7--6), [C\,\textsc{i}], and [C\,\textsc{ii}], enabling multi-tracer $M_{\rm H_2}$ estimates. For single-tracer masses we adopt: $\delta_{\rm GDR}=100$ for dust; for CO(7--6), $r_{71}=0.17$ and $\alpha_{\rm CO}=0.8~M_\odot\,({\rm K\,km\,s^{-1}\,pc^2})^{-1}$; for [C\,\textsc{i}](2--1), $T_{\rm ex,[CI]}=47$\,K and $X_{\rm CI}=1.9\times10^{-5}$; for [C\,\textsc{ii}], $T_{\rm ex,[CII]}=100$\,K and $X_{\rm C^+}=3.4\times10^{-4}$ (equivalently $\alpha_{\rm [CII]}\!\approx\!14~M_\odot\,L_\odot^{-1}$). Under these fiducials, the single-tracer $M_{\rm H_2}$ values are consistent within $\lesssim$0.3--0.5\,dex (i.e., a factor $\lesssim 3$).

    \item Using a hierarchical Bayesian cross–calibration that couples all four tracers, we obtain self-consistent $M_{\rm H_2}$ per source and tight posteriors for the conversion parameters. Results from our 18-object subsample agree with those from the full $z\!\sim\!6$ quasar set, indicating stable calibrations (e.g., $\alpha_{\rm [CII]}\!\approx\!14~M_\odot\,L_\odot^{-1}$, ${\rm \delta_{\rm GDR}}\!\sim\!100$) and modest intrinsic scatters ($\sim$0.1–0.2\,dex). While [C\,\textsc{ii}] may underpredict masses in compact, dust-hot systems exhibiting [C\,\textsc{ii}] deficits, the joint posteriors mitigate such biases and provide the most precise, uniform gas–mass estimates across heterogeneous line coverage.
  
    \item Excitation diagnostics based on multiple line ratios reveal highly excited and diverse ISM conditions in the $z\sim6$ quasar hosts. Comparison of the observed $L_{\rm [CII]}/L_{\rm [CI]}$ and $L_{\rm CO(7\text{--}6)}/L_{\rm TIR}$ ratios with PDR and XDR model grids shows that only a subset of sources can be reproduced by pure PDR models, corresponding to typical gas densities of $n \gtrsim 10^4$~cm$^{-3}$ and strong radiation fields of $G_0 \sim 10^3$--$10^4$. We note that such PDR-based interpretations may not reflect the average conditions of the molecular ISM. Moreover, the majority of objects fall outside the parameter space covered by pure PDR or XDR models, indicating that galaxy-integrated ISM conditions in these systems cannot be fully described by a single heating mechanism. Empirically, relative to typical SMGs and (U)LIRGs, quasars occupy the regime of elevated $L_{\rm CO(7\text{--}6)}/L_{\rm IR}$, consistent with compact and strongly excited gas. In addition, the $L'_{\rm CO(7-6)}/L'_{\rm [CI](2-1)}$ ratios are systematically higher than those observed in local LIRGs, main-sequence galaxies, and most SMGs, implying that a large fraction of the molecular gas is in a warm and highly excited phase. Taken together, these diagnostics suggest that the observed line ratios cannot be explained by classical PDR heating alone. Additional volumetric processes such as X-ray irradiation, supernova-driven turbulence and shocks, and enhanced cosmic-ray heating likely also contribute to the excitation and thermal balance of the cold ISM.

    \item Combining dust, CO, [C\,\textsc{i}], and [C\,\textsc{ii}] measurements offers a powerful tool to probe ISM excitation and constrain the total molecular gas reservoir in early quasar host galaxies. These results demonstrate the importance of multi-line diagnostics in disentangling star formation and AGN-driven heating in the early Universe.
\end{itemize}

Disentangling the physical driver of the line--to--IR deficits in $z\!\sim\!6$ quasar hosts will require targeted spectroscopy beyond [C\,\textsc{ii}]. High--$S/N$ measurements of [N\,\textsc{ii}]\,122/205\,$\mu$m (ionized--gas fraction and radiation hardness), [O\,\textsc{i}]\,63\,$\mu$m and [O\,\textsc{iii}]\,88\,$\mu$m (cooling budget), and $^{13}$[C\,\textsc{ii}] or velocity--resolved [C\,\textsc{ii}] (optical depth/saturation) will directly test whether AGN--heated dust, enhanced ionization to C$^{2+}$, or alternative coolants dominate. Sub-kiloparsec ALMA maps of [C\,\textsc{ii}], [C\,\textsc{i}], and mid--$J$/low--$J$ CO, together with mid--IR AGN diagnostics (e.g., PAH suppression, high--ionization lines), can separate compact PDRs from XDRs and quantify AGN contributions to $L_{\rm TIR}$. Incorporating these diagnostics along with dynamical masses into our hierarchical cross--calibration framework will tighten global conversion factors and per--source $M_{\rm H_2}$, break degeneracies among $T_{\rm ex}$, $\kappa_\nu$, and ${\delta_{\rm GDR}}$, and establish whether the quasar--host environment, rather than PDR physics alone, drives the observed deficits at cosmic dawn.

\begin{acknowledgements}
We thank the referee, Prof. Padelis Papadopoulos, for constructive comments that improved the quality of this work. F.X., R.D., and M.C.~acknowledge support from the INAF GO 2022 grant ``The birth of the giants: JWST sheds light on the build-up of quasars at cosmic dawn'', INAF Minigrant 2024 ``The interstellar medium at high redshift'', and by the PRIN MUR ``2022935STW'', RFF M4.C2.1.1, CUP J53D23001570006 and C53D23000950006. F.X. and R.W. acknowledge support from the National Natural Science Foundation of China (NSFC) with grant Nos. 12173002, 11991052, and the Ministry of Science and Technology of the People's Republic of China with grant No. 2022YFA1602902. A.P. acknowledges support from Fondazione Cariplo grant no. 2020-0902, the European Research Council (ERC) Consolidator Grant 864361 (CosmicWeb), and the Independent Research Fund Denmark (DFF) under grant 3120-00043B. D.R. gratefully acknowledges support from the Collaborative Research Center 1601 (SFB 1601 sub-projects C1, C2, C3, and C6) funded by the Deutsche Forschungsgemeinschaft (DFG) – 500700252. This paper makes use of the following ALMA data: ADS/JAO.ALMA\#2019.1.00147.S. ALMA is a partnership of ESO (representing its member states), NSF (USA) and NINS (Japan), together with NRC (Canada), NSTC and ASIAA (Taiwan), and KASI (Republic of Korea), in cooperation with the Republic of Chile. The Joint ALMA Observatory is operated by ESO, AUI/NRAO and NAOJ.\end{acknowledgements}

\bibliographystyle{aa}
\bibliography{paper_co76_survey_v5}

\begin{appendix} 
\onecolumn
\section{Sample and observation properties\label{app:sample}}

\begin{table*}[htbp]
\caption{Quasar sample and observed [C\,\textsc{ii}] and TIR properties.}
\label{tab:cii_sample}
\centering
\begin{tabular}{cccccccccc}
\hline\hline
Name & R.A. & Decl. & z & $M_{1450}$ & $F_{\rm [CII]}$ & $FWHM_{\rm [CII]}$ & $F_{\rm cont}$(1 mm) & $L_{\rm TIR}$ & Ref\\
 & (J2000) & (J2000) &  & (mag) & (Jy km s$^{-1}$) & (km s$^{-1}$) & [mJy] &($10^{12}L_\odot$) & \\
 (1) & (2) & (3) & (4) & (5) & (6) & (7) & (8) &(9) & (10)\\
\hline
J1120+0641 & 11:20:01.480 & +06:41:24.30 & 7.0842 & -26.63 & $1.01 \pm 0.09$ & $416 \pm 39$ & $0.64 \pm 0.05$ & $1.88 \pm 0.15$ & 1,6\\
J1104+2134 & 11:04:21.580 & +21:34:28.90 & 6.7662 & -26.63 & $1.35 \pm 0.21$ & $664 \pm 95$ & $1.99 \pm 0.04$ & $5.71 \pm 0.11$ & 2,7\\
J1048-0109 & 10:48:19.086 & -01:09:40.29 & 6.6759 & -26.0 & $2.52 \pm 0.07$ & $330 \pm 33$ & $2.84 \pm 0.03$ & $7.54 \pm 0.08$ & 3,8\\
PJ231-20 & 15:26:37.837 & -20:50:00.75 & 6.5864 & -27.2 & $2.65 \pm 0.12$ & $404 \pm 39$ & $3.36 \pm 0.05$ & $8.68 \pm 0.13$ & 3,9\\
PJ167-13 & 11:10:33.979 & -13:29:45.82 & 6.5148 & -25.62 & $2.53 \pm 0.07$ & $437 \pm 34$ & $0.87 \pm 0.05$ & $2.32 \pm 0.13$ & 3,6\\
J2318-3113 & 23:18:18.351 & -31:13:46.35 & 6.4435 & -26.11 & $1.11 \pm 0.14$ & $234 \pm 49$ & $0.57 \pm 0.09$ & $1.43 \pm 0.23$ & 3,3\\
PJ159-02 & 10:36:54.191 & -02:32:37.94 & 6.3809 & -26.8 & $1.15 \pm 0.07$ & $373 \pm 40$ & $0.65 \pm 0.03$ & $1.64 \pm 0.08$ & 3,6\\
J2211-3206 & 22:11:12.391 & -32:06:12.94 & 6.3394 & -26.71 & $0.57 \pm 0.11$ & $529 \pm 118$ & $0.57 \pm 0.05$ & $1.43 \pm 0.13$ & 3,3\\
J0142-3327 & 01:42:43.730 & -33:27:45.47 & 6.3379 & -27.81 & $2.62 \pm 0.06$ & $300 \pm 32$ & $1.65 \pm 0.04$ & $4.05 \pm 0.10$ & 3,6\\
PJ308-21 & 20:32:10.000 & -21:14:02.30 & 6.2341 & -26.35 & $1.79 \pm 0.10$ & $570 \pm 45$ & $0.64 \pm 0.05$ & $1.53 \pm 0.12$ & 3,3\\
PJ065-26 & 04:21:38.052 & -26:57:15.60 & 6.1877 & -27.25 & $2.05 \pm 0.11$ & $517 \pm 44$ & $1.23 \pm 0.05$ & $2.87 \pm 0.12$ & 3,6\\
PJ359-06 & 23:56:32.451 & -06:22:59.26 & 6.1722 & -26.79 & $2.47 \pm 0.16$ & $330 \pm 39$ & $0.87 \pm 0.08$ & $2.03 \pm 0.19$ & 3,6\\
PJ217-16 & 14:28:21.394 & -16:02:43.29 & 6.1498 & -26.93 & $0.70 \pm 0.11$ & $491 \pm 75$ & $0.37 \pm 0.06$ & $0.88 \pm 0.14$ & 3,6\\
J2219+0102 & 23:18:33.100 & -30:29:33.37 & 6.1492 & -23.1 & $2.54 \pm 0.13$ & $264 \pm 15$ & $0.766 \pm 0.047$ & $1.76 \pm 0.11$ & 4,4\\
J2318-3029 & 13:19:11.302 & +09:50:51.49 & 6.1458 & -26.21 & $2.34 \pm 0.12$ & $320 \pm 34$ & $2.71 \pm 0.08$ & $6.60 \pm 0.19$ & 3,3\\
J1319+0950 & 04:22:00.994 & -19:27:28.68 & 6.133 & -27.05 & $4.34 \pm 0.60$ & $515 \pm 81$ & $5.23 \pm 0.10$ & $12.3 \pm 0.24$ & 5,6\\
PJ065-19 & 15:09:41.778 & -17:49:26.80 & 6.1247 & -26.62 & $0.69 \pm 0.08$ & $345 \pm 67$ & $0.46 \pm 0.05$ & $1.08 \pm 0.12$ & 3,6\\
J1509-1749 & 22:19:17.217 & +01:02:48.90 & 6.1225 & -27.14 & $1.50 \pm 0.12$ & $631 \pm 72$ & $1.72 \pm 0.05$ & $4.06 \pm 0.12$ & 3,6\\

\hline
\end{tabular}
\tablefoot{
(1) Source name; 
(2)–(3) J2000 coordinates; 
(4) Redshift derived from [C\,\textsc{ii}]; 
(5) Absolute UV magnitude; 
(6) [C\,\textsc{ii}] integrated flux; 
(7) [C\,\textsc{ii}] FWHM; 
(8) Continuum flux density at 1.2\,mm; 
(9) Total infrared luminosity with uncertainty; 
(10) References for the [C\,\textsc{ii}] and $M_{1450}$ measurements: 
1–\citet{Venemans2020}, 2–\citet{Wang2024}, 3–\citet{Decarli2018}, 4–\citet{Willott2017}, 5–\citet{Wang2013}, 6-\citet{Banados2016}, 7-\citet{Yang2021}, 8-\citet{wang2017}, 9-\citet{Connor2020}.
}
\end{table*}

\begin{table*}[htbp]
\centering
\caption{Observing log and noise rms for the [C\,\textsc{ii}]-selected quasar sample.\label{tab:obs}}
\begin{tabular}{lcccccc}
\hline\hline
Short Name & Date Obs. & Exp. Time & Beam & Beam PA & rms (CO(7--6)) & rms ([C\,\textsc{i}]) \\
             &           &  [min]     & [\arcsec]  & [deg]    & [mJy beam$^{-1}$] & [mJy beam$^{-1}$] \\
\hline
J1120+0641 & 2019-11-03 & 34.776 & $2.77 \times 2.62$ & 80.3 & 0.27 & 0.28 \\
J1104+2134 & 2019-11-05 & 40.824 & $2.31 \times 1.94$ & -32.8 & 0.18 & 0.18 \\
J1048-0109 & 2019-10-31 & 34.272 & $2.23 \times 1.64$ & 68.0 & 0.23 & 0.24 \\
PJ231-20 & 2019-10-06 & 33.768 & $1.02 \times 0.81$ & -85.8 & 0.21 & 0.23 \\
PJ167-13 & 2020-03-03 & 34.776 & $1.27 \times 1.09$ & -79.6 & 0.24 & 0.26 \\
J2318-3113 & 2019-10-20 & 35.280 & $1.44 \times 1.06$ & -78.9 & 0.21 & 0.23 \\
PJ159-02 & 2019-10-31 & 38.304 & $2.20 \times 1.46$ & 69.9 & 0.24 & 0.25 \\
J2211-3206 & 2019-10-13 & 37.800 & $1.28 \times 1.08$ & -78.4 & 0.20 & 0.21 \\
J0142-3327 & 2019-10-20 & 37.800 & $1.40 \times 1.07$ & -77.6 & 0.18 & 0.20 \\
PJ308-21 & 2019-10-14 & 41.832 & $1.26 \times 1.07$ & -71.4 & 0.18 & 0.20 \\
PJ065-26 & 2019-10-19 & 44.856 & $1.36 \times 1.03$ & -66.7 & 0.21 & 0.23 \\
PJ359-06 & 2019-10-14 & 47.880 & $1.22 \times 1.09$ & 87.2 & 0.20 & 0.21 \\
PJ217-16 & 2019-11-03 & 48.888 & $2.20 \times 1.71$ & 88.6 & 0.25 & 0.27 \\
J2219+0102 & 2019-10-20 & 52.416 & $1.27 \times 1.07$ & -50.9 & 0.21 & 0.23 \\
J2318-3029 & 2019-10-17 & 49.896 & $1.31 \times 1.13$ & -62.3 & 0.20 & 0.21 \\
J1319+0950 & 2019-11-02 & 60.480 & $2.00 \times 1.60$ & -61.3 & 0.24 & 0.25 \\
PJ065-19 & 2019-10-19 & 52.416 & $1.26 \times 1.02$ & -69.4 & 0.23 & 0.24 \\
J1509-1749 & 2019-11-10 & 52.416 & $2.11 \times 1.58$ & -82.0 & 0.24 & 0.26 \\
\hline
\end{tabular}
\tablefoot{
(1) Source name; (2) Observation Date; (3) Integration Time; (4)-(5) Beam (Major $\times$ Minor Axis, Position Angle), (6)–(7): Median rms values (in mJy beam$^{-1}$) measured near the CO(7–6) and [C \textsc{i}] lines, respectively, using 50 km s$^{-1}$ channel binning. For source J1104+2134,  80~km\,s$^{-1}$ was used to improve sensitivity.
}
\end{table*}

\section{Velocity-integrated maps and spectra of the full sample \label{app:figs}}

\begin{figure*}
    \centering
    \includegraphics[width=1\textwidth]{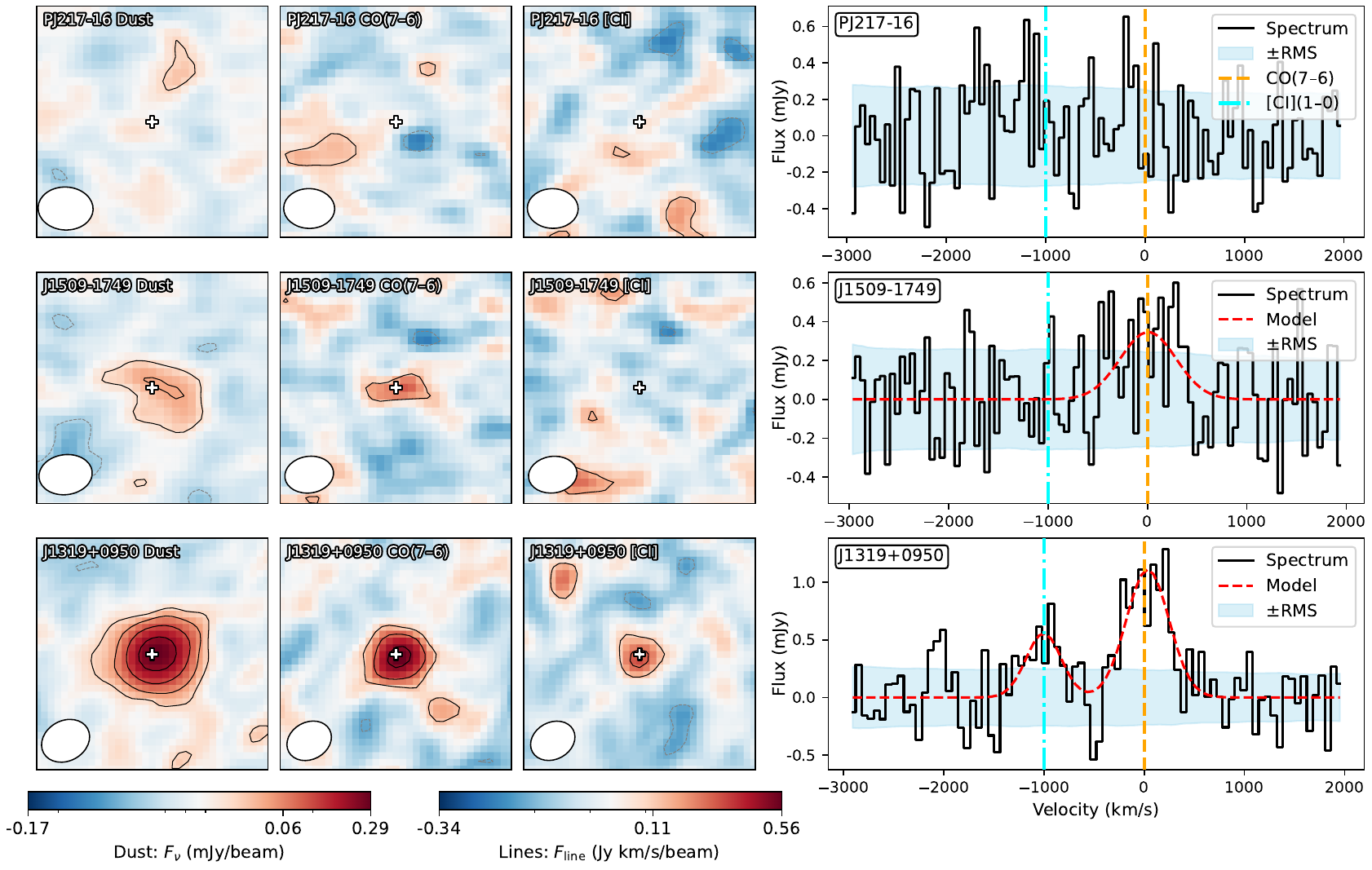}
    \caption{
    {\bf Left:} Velocity-integrated line and continuum maps for three quasars: PJ217--16, J1509--1749, and J1319+0950. Each row corresponds to one source, with panels (from left to right) showing the rest-frame FIR dust continuum, CO(7--6), and [C\,\textsc{i}](2--1) emission. Contours are shown at $[\pm 2, \pm 4, \pm 8, \pm 16]\,\sigma$, and the synthesized beam is shown in the lower-left corner. The cross marks the optical/NIR centroid.
    {\bf Right:} Corresponding CO(7--6) spectra extracted at the source positions. The red dashed line shows single-Gaussian fits to each detected line, and the shaded region indicates the 1$\sigma$ noise level.
    }
    \label{fig:group1}
\end{figure*}

\begin{figure*}[htbp]
    \centering
    \includegraphics[width=0.95\textwidth]{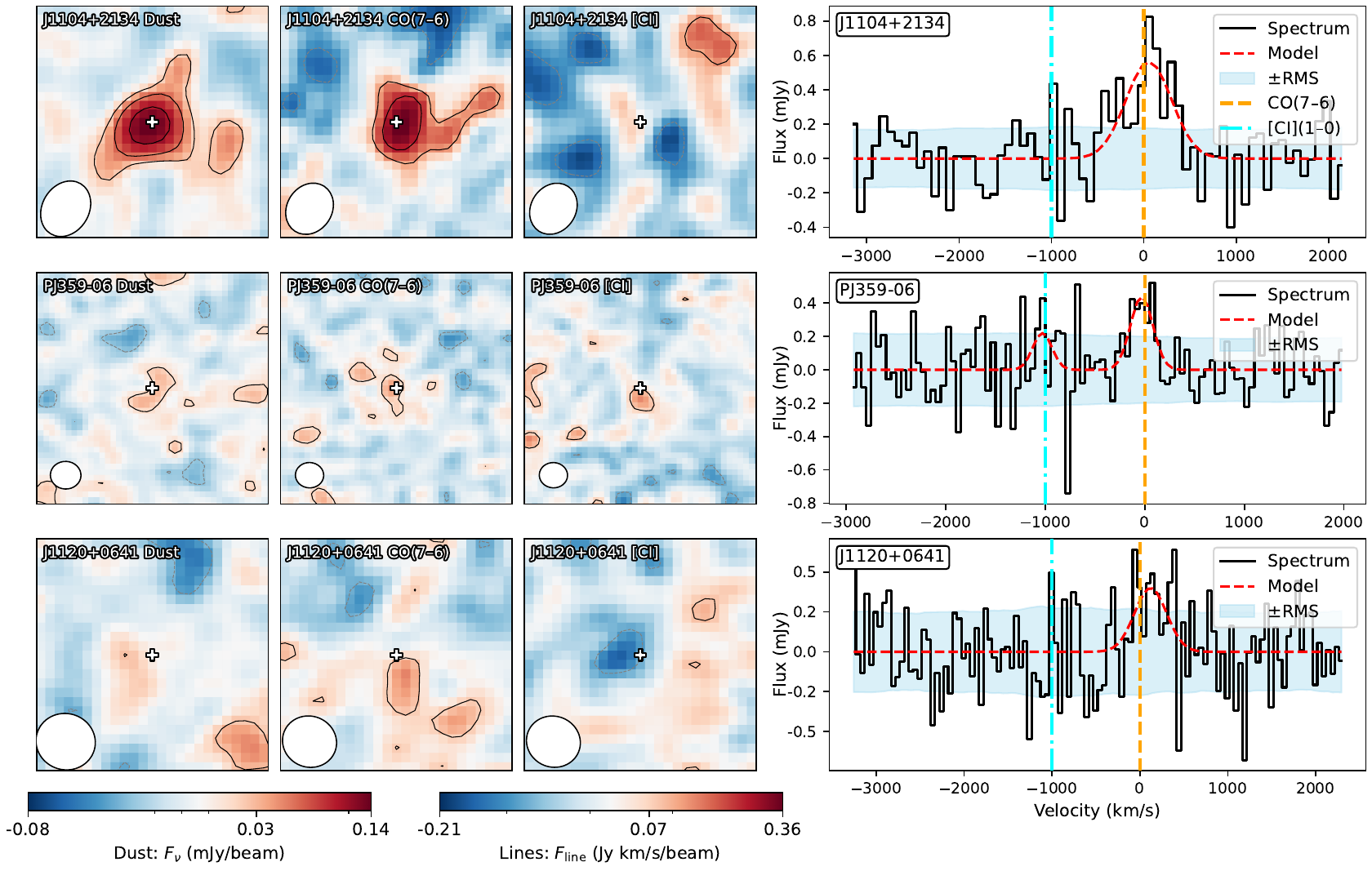}
    \caption{
    Same quantities and layout as in Fig.~\ref{fig:group1}, but for three quasars (from top to bottom): J1104+2134, PJ359-06, and J1120+0641.
    }
    \label{fig:group2}
\end{figure*}

\begin{figure*}[htbp]
    \centering
    \includegraphics[width=0.95\textwidth]{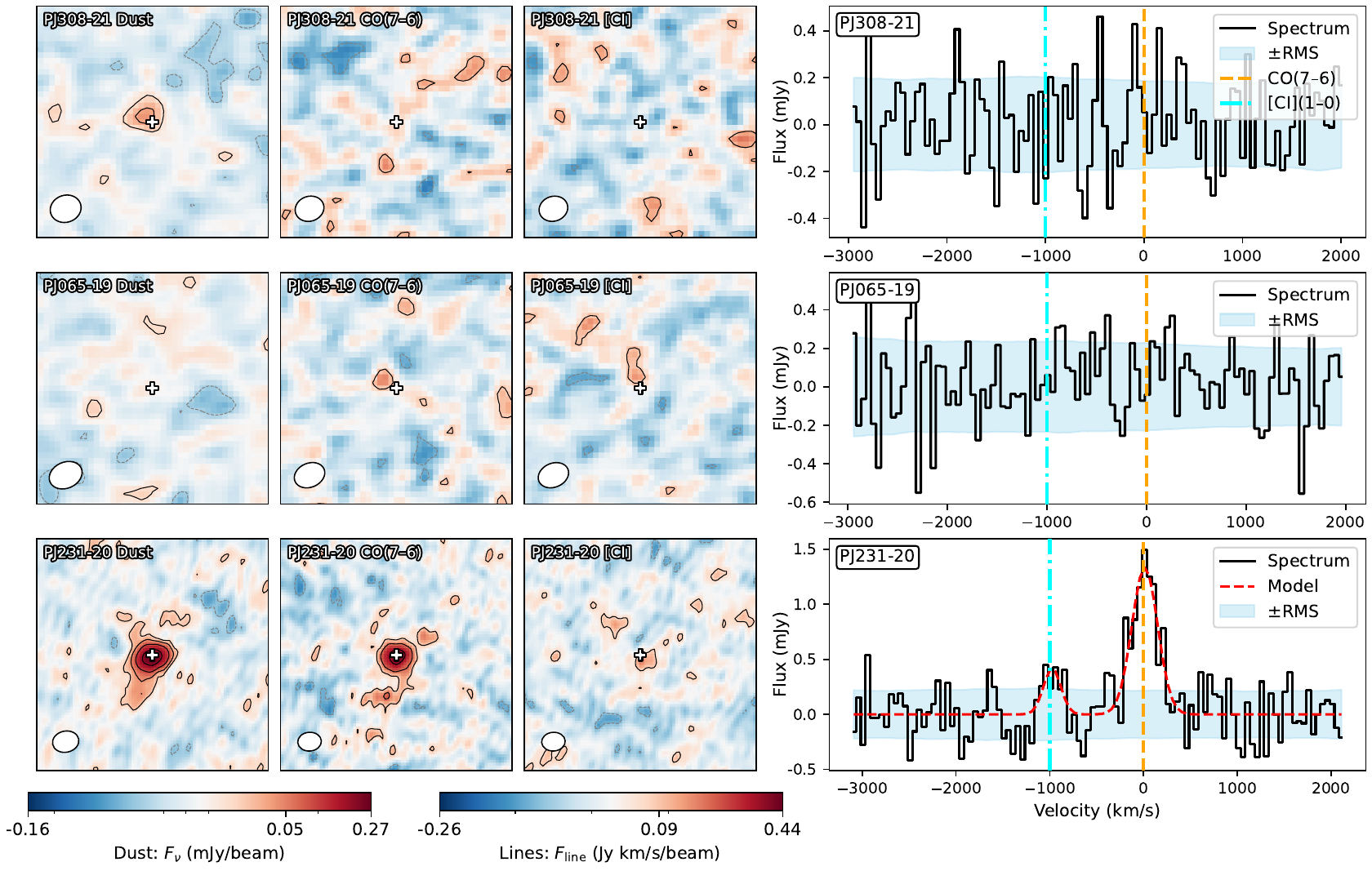}
    \caption{
    Same quantities and layout as in Fig.~\ref{fig:group1}, but for three quasars (from top to bottom): PJ308-21, PJ065-19, and PJ231-20.
    }
    \label{fig:group3}
\end{figure*}

\begin{figure*}[htbp]
    \centering
    \includegraphics[width=0.95\textwidth]{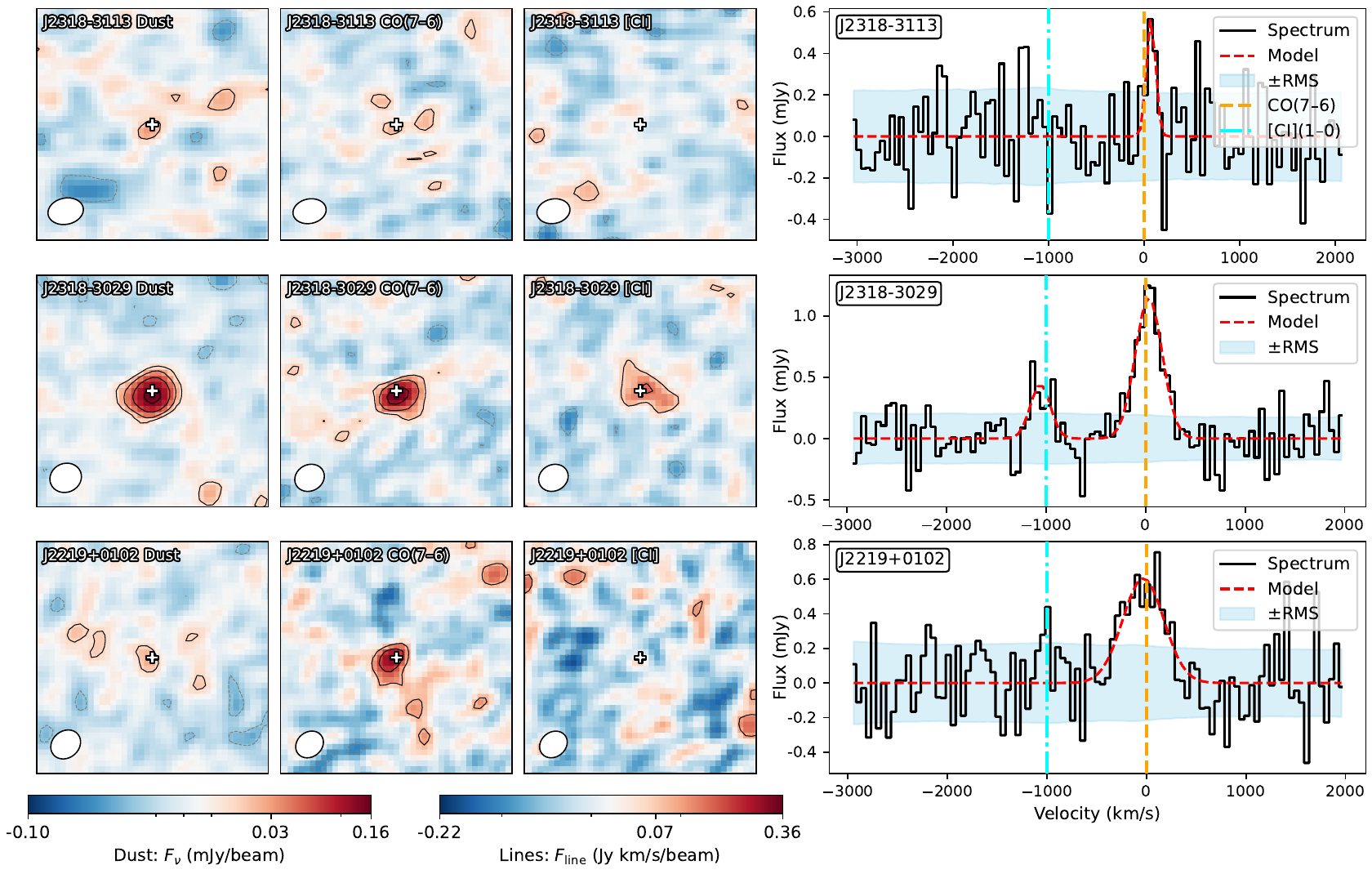}
    \caption{
    Same quantities and layout as in Fig.~\ref{fig:group1}, but for three quasars (from top to bottom): J2318-3113, J2318-3029, and J2219+0102.
    }
    \label{fig:group4}
\end{figure*}

\begin{figure*}[htbp]
    \centering
    \includegraphics[width=0.95\textwidth]{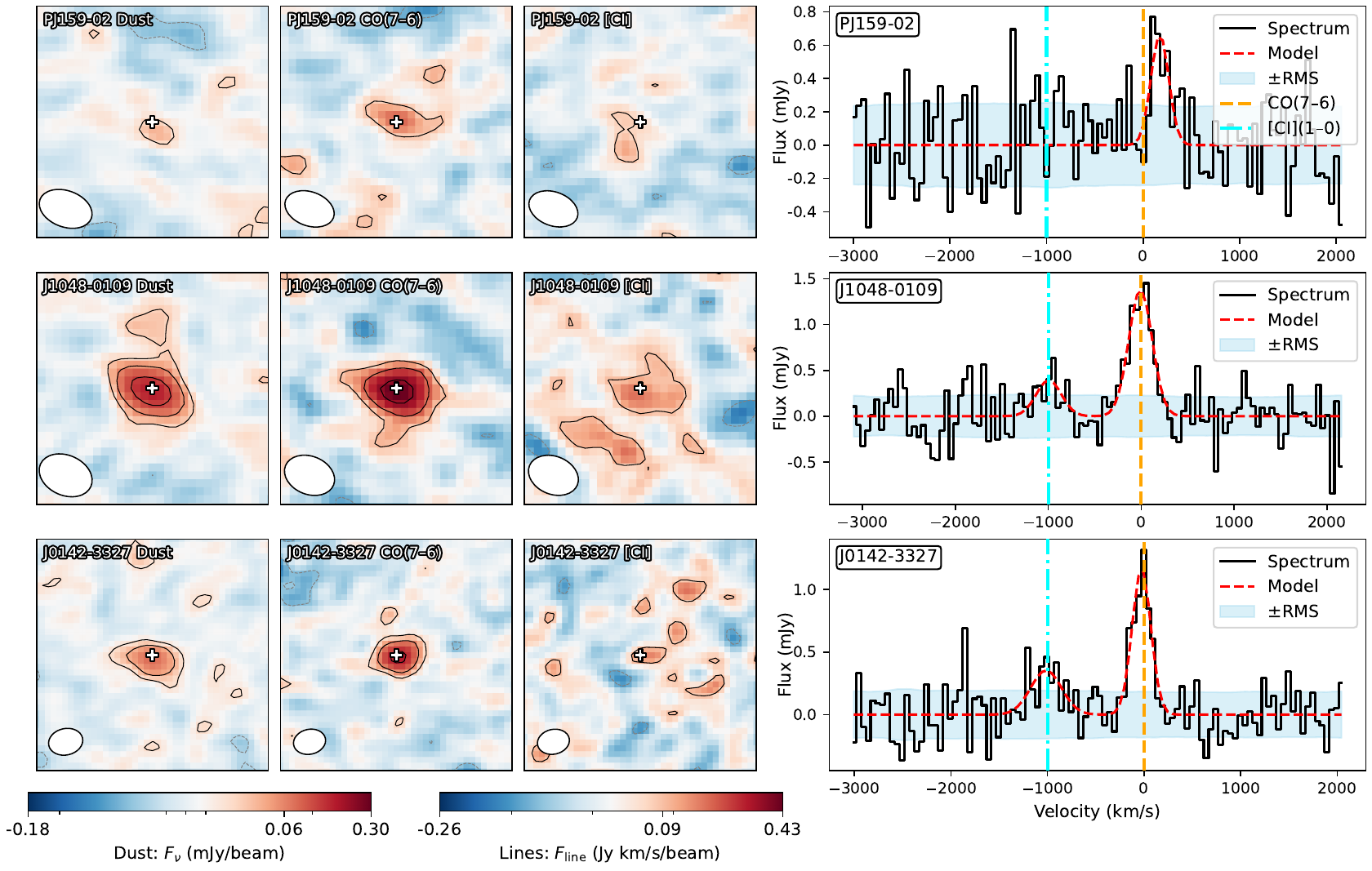}
    \caption{
    Same quantities and layout as in Fig.~\ref{fig:group1}, but for three quasars (from top to bottom): PJ159-02, J1048-0109, and J0142-3327.
    }
    \label{fig:group5}
\end{figure*}

\begin{figure*}[htbp]
    \centering
    \includegraphics[width=0.95\textwidth]{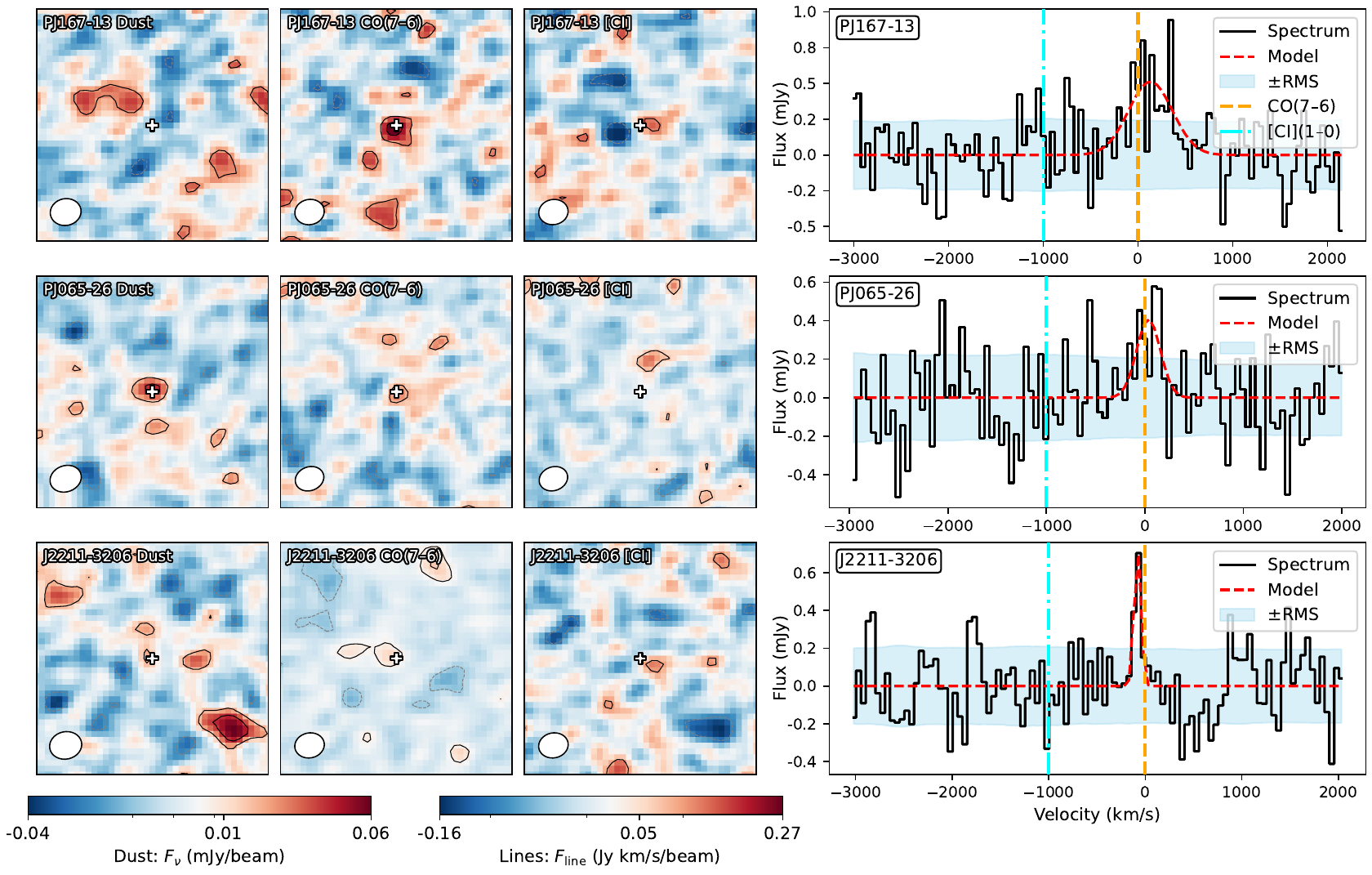}
    \caption{
    Same quantities and layout as in Fig.~\ref{fig:group1}, but for three quasars (from top to bottom): PJ167-13, PJ065-26, and J2211-3206.
    }
    \label{fig:group6}
\end{figure*}

Figures~\ref{fig:group1}--\ref{fig:group6} present the velocity-integrated line and continuum maps for the remaining 18 quasars in our sample (three sources per figure). Each figure shows, in the left panel, the velocity-integrated dust continuum, CO(7--6), and [C\,\textsc{i}](2--1) maps, with contours at $[\pm2, \pm4, \pm8, \dots]\,\sigma$ and the
synthesized beam in the lower-left corner; and in the right panel, the
corresponding CO(7--6) spectra extracted at the source positions. As in
Fig.~\ref{fig:group1}, the red dashed line shows a Gaussian fit to the
line profile, and the shaded region indicates the $1\sigma$ noise level.

In total, these five figures cover the 15 quasars not shown in the main
text, thereby completing the presentation of our full sample of 18
sources.

\FloatBarrier

\section{Supplementary continuum and line luminosity ratios \label{app:dust3to1}}

\begin{figure}[htbp]
  \centering
  \includegraphics[width=0.5\linewidth]{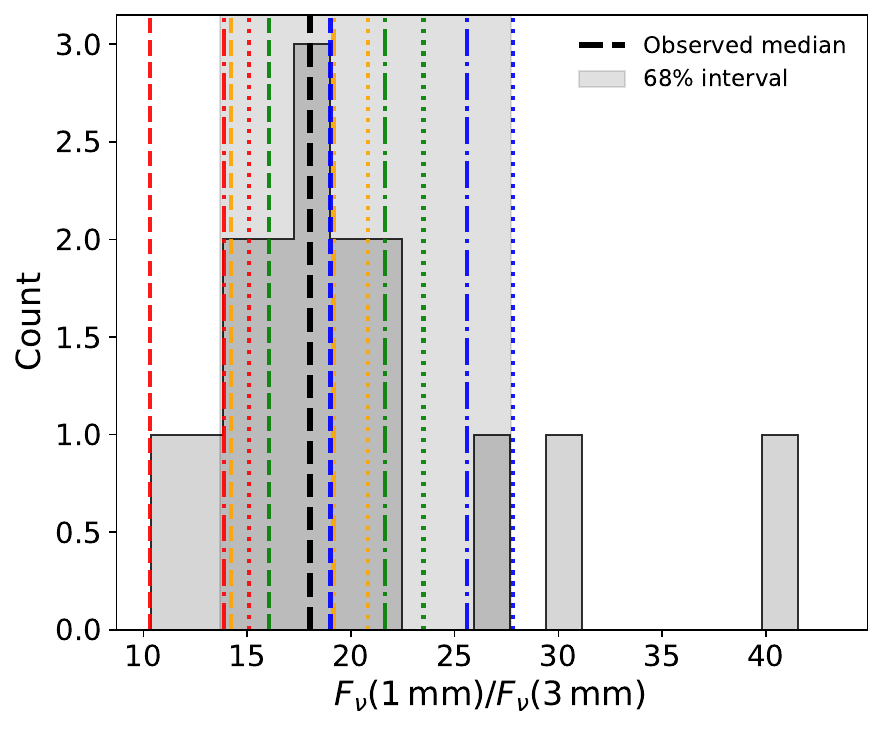}
  \caption{Distribution of the observed continuum flux–density ratio
  $R \equiv F_{\nu}(1\,\mathrm{mm})/F_{\nu}(3\,\mathrm{mm})$.
  The dashed line marks the median ratio; the shaded band indicates the central 68\% interval (16th–84th percentiles).
  Model expectations for selected $(T_{\rm d},\tau)$ are overplotted as vertical lines, using the same color/linestyle convention as Figure ~\ref{fig:mm-cont}.}
  \label{fig:ratio-hist}
\end{figure}

\twocolumn
\begin{figure}[htbp]
    \centering
    \includegraphics[width=\linewidth]{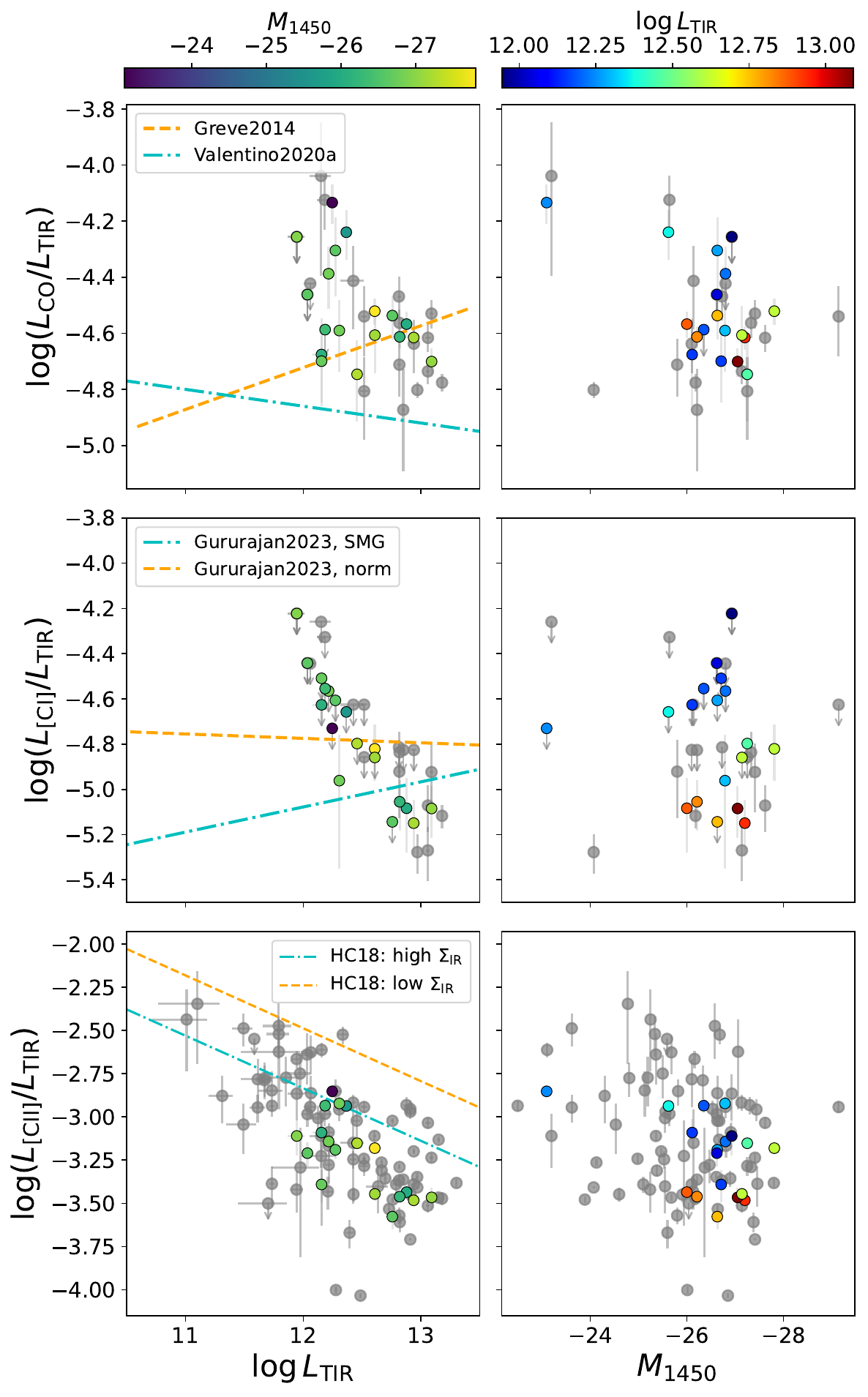}
    \caption{Similar to Figure \ref{fig:luminosity_comparison}, but the y-axis shows the line-to-TIR luminosity ratio, $L_{\rm line}/L_{\rm TIR}$. Rows (top to bottom): CO(7--6), [C\,\textsc{i}], and [C\,\textsc{ii}].  
    }
    \label{fig:lum_linetoTIR}
\end{figure}

\begin{figure}[htbp]
  \centering
  \includegraphics[width=\linewidth]{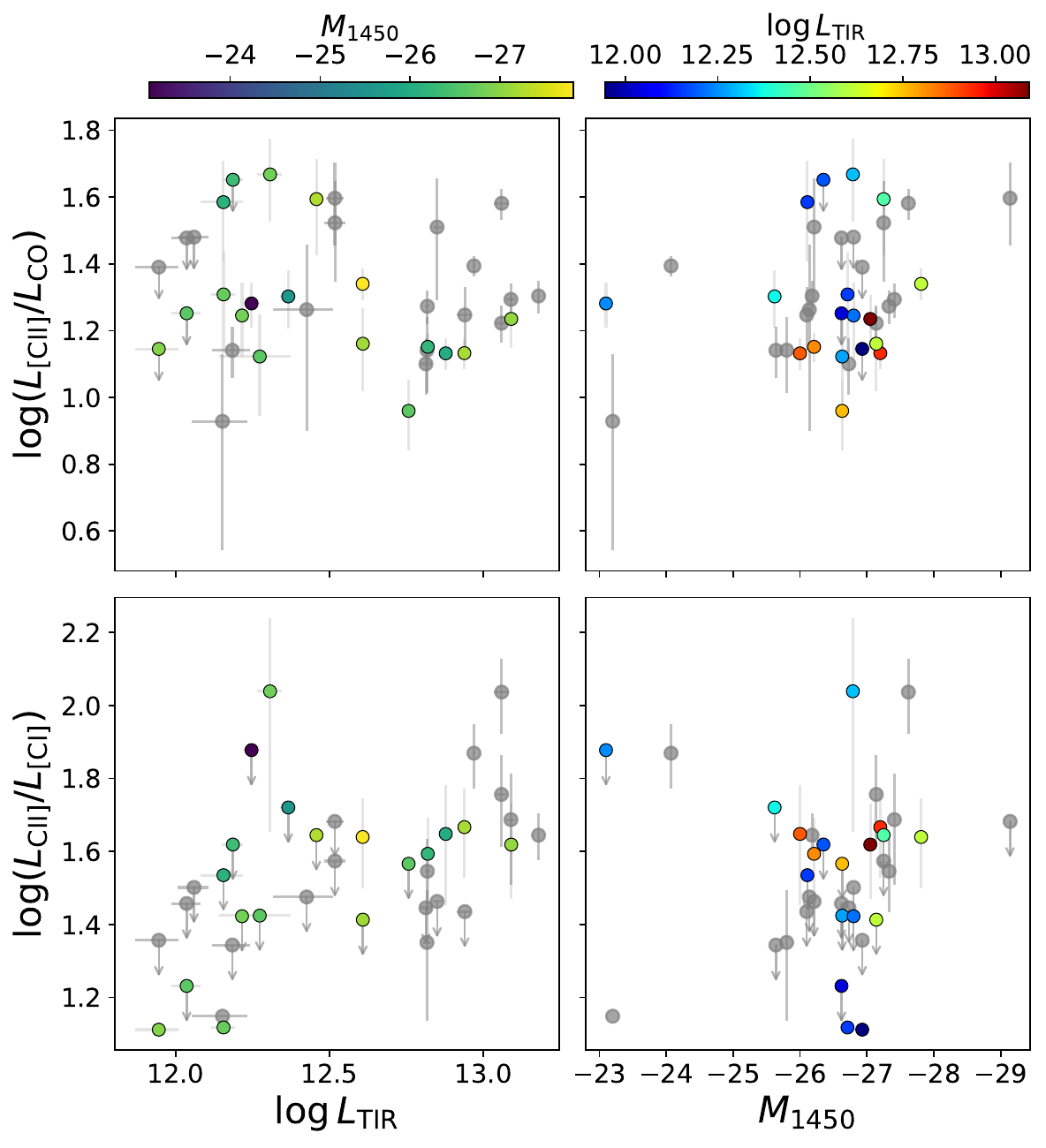}
  \caption{Similar to Figure~\ref{fig:luminosity_comparison}, but the y-axis shows the [C\,\textsc{ii}]--to--line luminosity ratio, $L_{\rm [CII]}/L_{\rm line}$. 
  Rows (top to bottom): $L_{\rm [CII]}/L_{\rm CO(7\text{--}6)}$ and $L_{\rm [CII]}/L_{\rm [CI]}$. 
  Error bars and upper limits are shown where available; other plotting styles follow Fig.~\ref{fig:luminosity_comparison}.}
  \label{fig:lum_ciitoline}
\end{figure}

\onecolumn
\section{Supplementary results of the hierarchical cross–calibration \label{app:cross}}

\begin{figure*}[htbp]
  \centering
  \includegraphics[width=0.9\linewidth]{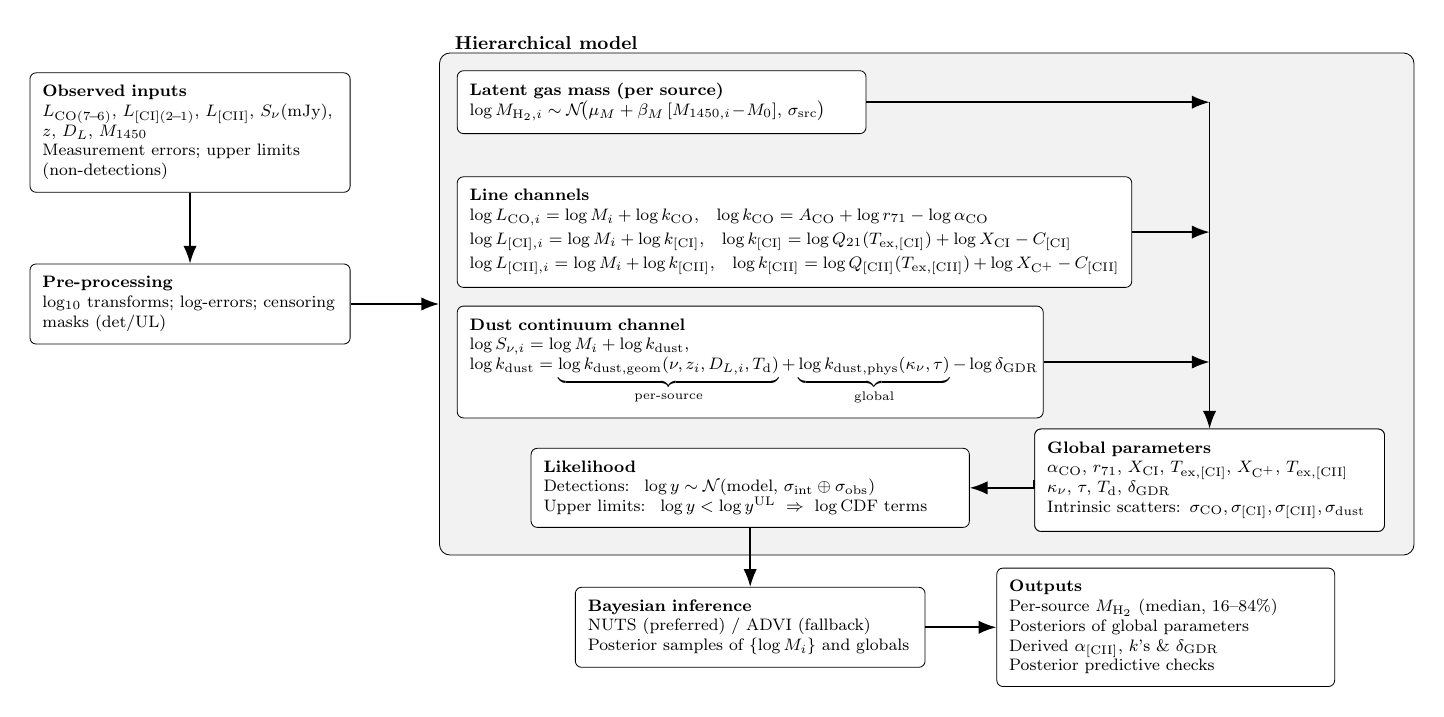}
  \caption{Concise workflow of the hierarchical Bayesian cross–calibration used in this work. 
  Inputs (measured line luminosities, continuum fluxes and uncertainties, with upper limits) are mapped into a hierarchical model where each tracer obeys 
  $\log L = \log M_{\rm H_2}+\log k_{\rm tracer}$ (and $\log S_\nu$ for dust), with physically–motivated global parameters.
  Censored data enter via CDF likelihood terms. We infer per–source $M_{\rm H_2}$ and global coefficients using NUTS (or ADVI).}
  \label{fig:bayes_flow}
\end{figure*}

\begin{table*}[!htbp]
\centering
\caption{Supplementary results of the hierarchical cross–calibration.}
\label{tab:xcal_post}
\begin{tabular}{lccc}
\hline
Parameter & Units & Our sample & All quasars \\
\hline
\multicolumn{4}{c}{k–aggregates}\\
$\log k_{\rm CO}$                               & —                    & $-2.46\;[-2.66,\,-2.25]$ & $-2.48\;[-2.69,\,-2.25]$ \\
$\log k_{\rm [CI]}$                             & —                    & $-2.98\;[-3.16,\,-2.74]$ & $-2.94\;[-3.14,\,-2.70]$ \\
$\log k_{\rm [CII]}$                            & —                    & $-1.13\;[-1.34,\,-0.93]$ & $-1.14\;[-1.35,\,-0.91]$ \\
$\log k_{\rm dust,\,phys}$                      & —                    & $60.24\;[60.08,\,60.41]$ & $60.25\;[60.04,\,60.45]$ \\
$\log k_{\rm dust,\,phys}-\log{\delta_{\rm GDR}}$ & —                  & $58.23\;[57.99,\,58.44]$ & $58.17\;[57.93,\,58.42]$ \\
\hline
\multicolumn{4}{c}{Line–physics parameters}\\
$\alpha_{\rm CO}$          & $(\mathrm{K\,km\,s^{-1}\,pc^2})^{-1}$     & $0.81\;[0.37,\,1.53]$    & $0.80\;[0.40,\,1.62]$ \\
$r_{71}$                                        & —                    & $0.18\;[0.09,\,0.32]$    & $0.17\;[0.09,\,0.32]$ \\
$T_{\rm ex,[CI]}$                               & K                    & $48.67\;[38.81,\,55.99]$ & $46.84\;[37.45,\,56.59]$ \\
$T_{\rm ex,[CII]}$                              & K                    & $93.50\;[79.58,\,116.62]$  & $100.73\;[81.32,\,119.86]$ \\
$\log X_{\rm CI}$                               & —                    & $-4.81\;[-4.98,\,-4.55]$ & $-4.74\;[-4.96,\,-4.51]$ \\
$\log X_{\rm C^+}$                              & —                    & $-4.37\;[-4.61,\,-4.20]$ & $-4.40\;[-4.62,\,-4.17]$ \\
$\alpha_{\rm [CII]}$              &  $M_\odot\,L_\odot^{-1} $          & $13.46\;[8.55,\,21.73]$  & $13.71\;[8.07,\,22.28]$ \\
\hline
\multicolumn{4}{c}{Dust–physics parameters}\\
$\kappa_\nu$                                    & cm$^2$ g$^{-1}$      & $10.05\;[6.95,\,14.70]$  & $10.08\;[6.32,\,15.89]$ \\
$\tau$                                          & —                    & $0.23\;[0.11,\,0.44]$    & $0.20\;[0.10,\,0.40]$ \\
$T_{\rm d}$                                     & K                    & $45.59\;[39.00,\,54.31]$ & $45.42\;[37.33,\,54.56]$ \\
${\delta_{\rm GDR}}$                            & —                    & $102.09\;[70.31,\,172.58]$ & $119.67\;[68.55,\,204.64]$ \\
\hline
\multicolumn{4}{c}{Intrinsic scatters}\\
$\sigma_{\rm CO,int}$                           & dex                  & $0.13\;[0.08,\,0.18]$    & $0.08\;[0.04,\,0.12]$ \\
$\sigma_{\rm [CI],int}$                         & dex                  & $0.09\;[0.03,\,0.18]$    & $0.06\;[0.02,\,0.11]$ \\
$\sigma_{\rm [CII],int}$                        & dex                  & $0.18\;[0.13,\,0.25]$    & $0.21\;[0.18,\,0.25]$ \\
$\sigma_{\rm dust,int}$                         & dex                  & $0.12\;[0.02,\,0.17]$    & $0.22\;[0.19,\,0.26]$ \\
\hline
\multicolumn{4}{c}{Population–level mass prior (hyper-parameters)}\\
$\mu_M$                                         & dex                  & $10.35\;[10.15,\,10.57]$ & $10.36\;[10.12,\,10.57]$ \\
$\beta_M$                                       & dex\,mag$^{-1}$      & $-0.03\;[-0.09,\,0.05]$  & $-0.12\;[-0.16,\,-0.08]$ \\
$\sigma_{\rm src}$                              & dex                  & $0.30\;[0.26,\,0.36]$    & $0.39\;[0.36,\,0.43]$ \\
\hline
\end{tabular}
\\[2pt]
\tablefoot{Columns show results from our 18-object subsample and from the full $z\!\sim\!6$ quasar sample.}
\end{table*}

\end{appendix}

\end{document}